\makeatletter
\declare@file@substitution{revtex4-1.cls}{revtex4-2.cls}
\makeatother

\documentclass[twocolumn]{aastex631}

\received{}
\revised{}
\accepted{}

\submitjournal{The Astrophysical Journal}

\shorttitle{Merger Shock Surfaces in Merging Galaxy Clusters}
\shortauthors{Lee, Ryu, \& Kang.}

\begin{document}

\title{Morphology and Mach Number Distribution of Merger Shock Surfaces in Merging Galaxy Clusters}
\author[0000-0003-0439-3019]{Eunyu Lee}
\affiliation{Department of Physics, College of Natural Sciences, UNIST, Ulsan 44919, Korea}
\author[0000-0002-5455-2957]{Dongsu Ryu}
\affiliation{Department of Physics, College of Natural Sciences, UNIST, Ulsan 44919, Korea}
\author[0000-0002-4674-5687]{Hyesung Kang}
\affiliation{Department of Earth Sciences, Pusan National University, Busan 46241, Korea}
\correspondingauthor{Dongsu Ryu}\email{dsryu@unist.ac.kr}
\correspondingauthor{Hyesung Kang}\email{hskang@pusan.ac.kr}

\begin{abstract}


In a binary merger of two subclusters with comparable masses, a pair of merger shocks are typically generated, often manifesting as double radio relics. Using cosmological hydrodynamic simulations, we identify major merger events with mass ratio $1<\mathcal{M}_1/\mathcal{M}_2\lesssim4$ and impact parameter $b/r_{\rm vir,1}\lesssim1$, where $r_{\rm vir,1}$ is the virial radius of the heavier subcluster. We analyze merger shock surfaces approximately 1 Gyr after the pericenter passage, focusing on their morphology and the distribution of the Mach number, $M_s$, of their constituent shock zones. The shock surfaces exhibit an elongated shape with a minor-to-major axis ratio of $\sim0.6-0.9$ and cover the area of $\sim5-20\%$ of the enclosed sphere. The area ratio of the two shock surfaces roughly scales as $A_{\rm ss,2}/A_{\rm ss,1}\propto\mathcal{M}_1/\mathcal{M}_2$, typically positioning the larger shock ahead of the lighter subcluster. The axis connecting the two subclusters generally does not pass through the centers of the shock surfaces, due to the nonzero impact parameter and ambient turbulent flows. The distribution of $M_s$ of shock zones on each surface can be approximated by a log-normal function, peaking at $M_{s,\rm{peak}}\approx2-4.5$ and extending up to $\sim10$. The surface-area-weighted and X-ray-emissivity-weighted average Mach numbers are comparable, with ${\langle{M_s}\rangle}_{\rm{area}}\approx2.3-4.4$ and ${\langle{M_s}\rangle}_{X}\approx2-4$. In contrast, the cosmic-ray-energy-flux-weighted average Mach numbers are higher with ${\langle{M_s}\rangle}_{\rm{CR}}\approx3-5$. This discrepancy aligns with the differences between Mach numbers derived from X-ray and radio observations of radio relic shocks. On the other hand, we find that mostly ${\langle{M_s}\rangle}_{X}\gtrsim2$ for simulated merger shocks, although shocks with $M_{\rm X-ray}\lesssim2$ are often reported in observations.

\end{abstract}


\section{Introduction}\label{s1}

During the hierarchical formation of the large-scale structure (LSS) of the universe, galaxy clusters grow through the successive mergers of small subclusters and groups of galaxies. Such mergers are known to induce shock waves and turbulent flow motions in the intracluster medium (ICM). Energy dissipation via shocks and turbulence plays key roles not only in heating the ICM but also in amplifying magnetic fields and accelerating cosmic rays (CRs) \citep[e.g.,][]{sarazin2002,ryu2003,Ryu2008,brunetti&jones2014,porter2015,Roh2019,Wittor2021}.

In particular, binary mergers between two subclumps of comparable masses with small impact parameters are classified as major mergers and induce a pair of so-called ``merger shocks'' after the pericenter passage of dark matter (DM) cores \citep[e.g.,][]{Markevitch2007,Ha2018}. The overall picture of merger shocks has been extensively studied through numerical simulations. In simulations of idealized binary mergers with subclusters in hydrostatic equilibrium, it was shown that merger shocks exhibit different properties depending on the merger parameters, such as the mass ratio, impact parameter, viewing angle, etc \citep[e.g.,][]{Gabich&Blasi2003,Springel&Farrar2007,vanWeeren2011a,ZuHone2011, Molnar2017}.

The properties of merger shocks were also described in the context of the LSS formation \citep[e.g.,][]{Paul2011, Schmidt2017,Ha2018,lokas2023,LeeWK2024,Finner2024}. For instance, analyzing merging clusters found in cosmological hydrodynamic simulations, {\citet{Ha2018} described quantities, such as the time variations of the shock speed, $v_s$, the shock Mach number, $M_s$, and the energy dissipation at shock interfaces, in the inherently turbulent ICM}\footnote{As in our previous papers, $M_s$ denotes the shock Mach number, while $\mathcal{M}_1$ and $\mathcal{M}_2$ represent the masses of two subclusters.}. While a variety of mergers with diverse merger parameters are present in these simulations, \citet{Ha2018} concentrated on the cases of nearly head-on collision with the mass ratio of $\mathcal{M}_1/\mathcal{M}_2\approx 2$, replicating the characteristics of the well-studied Sausage relic in the cluster CIZA J2242.8+5301 \citep[e.g.,][]{vanWeeren2010,Hoang2017}. As illustrated in Figure 1 of \citet{Ha2018}, as two subclusters approach and gas clumps are compressed, ``equatorial shocks'' first appear in the equatorial plane, propagating toward the direction perpendicular to the merger axis. After the DM core passage, two ``axial shocks” launch along the merger axis in front of subclusters \citep[see also Figure 1 of][]{vanWeeren2011a}.

Merger shocks often manifest as double radio relics in the outskirts of merging clusters \citep[e.g.,][]{vanWeeren2011a, Hoang2018, Golovich2019}. With observations of around a hundred cases to date, it is widely accepted that radio relics exhibit diffuse synchrotron radiation emitted by CR electrons accelerated at merger shocks \citep[see][for reviews]{Feretti2012,bruggen2012,brunetti&jones2014,vanWeeren2019}. Although it has been customary to associate a merger shock with a single Mach number, considering the curved, or often irregular, morphology of radio relics and the turbulent nature of the ICM, the correct picture should be that a merger shock surface is composed of many ``shock zones'' characterized by different $M_s$ \citep[e.g.,][]{Ha2018,Roh2019,Botteon2020,Wittor2021}. Thus, the Mach number estimated from radio or X-ray observations should represent mean values, ${\langle M_s\rangle}_{\rm radio}$ and ${\langle M_s\rangle}_{\rm X-ray}$, for an ensemble of shock zones associated with each merger shock surface \citep[e.g.,][]{Hong2015,Rajpurohit2020, Dominguez-Fernandez2021}.

In radio observations, radio relics frequently appear as arc-shaped patches at a distance of $d_s\sim 1-2$ Mpc from the cluster center, usually around the virial radius, $r_{\rm vir}$. The Mach numbers of radio relic shocks are estimated using either the ``injection spectral index'', $\alpha_{\rm inj}$ (or $\alpha_{\rm sh}$), of the radio spectrum immediately behind the shock in the outer edge of observed relics, i.e., $ M_{\rm radio}\approx[(3+2\alpha_{\rm inj})/(2\alpha_{\rm inj}-1)]^{1/2}$, or the ``integrated spectral index'', $\alpha_{\rm int}$, of the volume-integrated spectrum, i.e., $M_{\rm radio}\approx[(\alpha_{\rm int}+1)/(\alpha_{\rm int}-1)]^{1/2}$ \citep[e.g.,][]{Kang2011}. Radio observations indicate typically $M_{\rm radio} \approx 2-5$ \citep{vanWeeren2019}. On the other hand, in X-ray observations, merger shocks are detected as discontinuities in the surface brightness or temperature profiles that can be used to estimate the Mach number, $M_{\rm X-ray}$ \citep[e.g.,][]{Markevitch2002, Ogrean2013, Dasadia2016, Sanders2022}. These shocks are typically found at $d_s \gtrsim$ 1 Mpc and mostly weak with $M_{\rm X-ray} \sim 1.2-4$ \citep[e.g.,][]{Eckert2016,Urdampilleta2018}.

For many observed radio relics, the Mach numbers derived from radio and X-ray observations do not coincide, exhibiting the so-called Mach number discrepancy \citep[e.g.,][]{Wittor2021}. In most cases, Mach numbers inferred from X-ray observations are lower than those inferred from radio observations, i.e., $M_{\rm X-ray}\lesssim M_{\rm radio}$ \citep[e.g.,][]{Akamatsu2013,vanWeeren2016}. For instance, in the so-called Toothbrush relic in 1RXS J0603.3, $M_{\rm radio} \approx 3.3$ with $\alpha_{\rm sh}\approx 0.7$ and $M_{\rm radio} \approx 3.78$ with $\alpha_{\rm int}\approx 1.15$ \citep{Rajpurohit2018}, whereas $M_{\rm X-ray} \lesssim 2$ \citep{Itahana2015,vanWeeren2016}. However, in some relics, the opposite is found; e.g., for a radio relic in Abell 521, $M_{\rm radio} \approx 2.27$ and $M_{\rm X-ray} \approx 3.4$ are reported \citep{Giacintucci2006}.

As noted above, a merger shock is a surface comprising many shock zones with varying Mach numbers. Previous studies have suggested that radio observations tend to pick up regions with high CR acceleration and, consequently, higher $M_s$, while X-ray observations preferentially reveal regions with higher densities and thus higher X-ray emissivity, which are typically associated with lower $M_s$ \citep[e.g.,][]{Hong2015,Ha2018,Roh2019,Botteon2020}. This proposition could explain why $M_{\rm radio}\gtrsim M_{\rm X-ray}$. However, complex flows surrounding merger shocks in the turbulent ICM could lead to the opposite case. These highlight the needs to understand the detailed properties of merger shocks and the dynamics of surrounding flows and to clarify the nature of observed radio relics.

In this work, we study the morphological characteristics of merger shock surfaces and the Mach number distribution of shock zones in major binary merging clusters with different mass ratios and impact parameters using cosmological hydrodynamic simulation data. To achieve this, we perform a set of structure formation simulations, identify synthetic merging clusters formed through major binary mergers, and track the time evolution of merger events such as the pericenter passage and the development of merger shocks. Using specific criteria, we try to isolate the surfaces of merger shocks from shocks generated by turbulence in the background ICM and the infall of the warm-hot intergalactic medium (WHIM) along filaments. We then analyze the shape, size, and orientation of the shock surfaces, and examine the probability distribution function (PDF) of $M_s$ for shock zones on these surfaces. Additionally, we estimate the average Mach numbers weighted by shock surface area, X-ray emissivity, and CR energy flux, and discuss the implications for observations of radio relics.

The paper is organized as follows: In Section \ref{s2}, we outline the numerical setup, including the generation of synthetic merging clusters and the method for identifying merger shock surfaces. Section \ref{s3} presents the results, discussing the morphological characteristics of merger shock surfaces and the Mach number distribution of shock zones. In Section \ref{s4}, we compare the average Mach numbers of merger shocks with those observed in radio relic shocks. Finally, a brief summary is provided in Section \ref{s5}.

\section{Numerics}\label{s2}

\subsection{Simulation Setup}\label{s2.1}

We generate samples of merging clusters and merger shocks from simulation data for the LSS formation of the universe. Simulations are performed using the particle-mesh/Eulerian cosmological hydrodynamic code described in \citet{Ryu1993}. We adopt the standard $\Lambda$CDM cosmology model with the following parameters: baryon density {$\Omega_b$ = 0.046}, dark matter density $\Omega_{\rm DM}$= 0.234, cosmological constant $\Omega_\Lambda$ = 0.72, Hubble parameter $h\equiv H_0$/(100 km s\(^{-1}\)Mpc\(^{-1}\)) = 0.7, primordial spectral index $n$ = 0.96, and rms density fluctuation $\sigma_8$ = 0.82, complying with \textit{WMAP}9 data \citep{Hinshaw2013}. A cubic box of the comoving size of $100~h^{-1}$Mpc with periodic boundaries is employed and divided into $2048^3$ uniform grid zones. This results in a uniform spatial grid resolution of $\Delta l$ = 48.8 $h^{-1}$ kpc. 

In our simulations, {non-gravitational processes, such as radiative cooling and galaxy formation feedback, are not included}, as their effects on the dynamical evolution of the ICM on megaparsec scales are expected to be insignificant. {Magnetic fields are also not included; given the expected magnetic field strength of $\lesssim 1 \mu$G at cluster outskirts \citep[e.g.,][]{Ryu2008,vazza2014}, the magnetic energy would be an order of magnitude smaller than the kinetic energy, and hence magnetic fields are unlikely to significantly affect the properties of merger shocks, such as their morphology and the Mach number distribution at the shock surface. On the other hand, on microscopic scales, the behavior of magnetic fields, such as amplification and dissipation through kinetic instabilities and wave-particle interactions, should play a crucial role in particle acceleration and subsequent synchrotron emission behind the shock \citep[e.g.,][]{Kang2019,Ha2021,Ha2022}, but such processes are beyond the scope of this study.}

\subsection{Merging Cluster Sample}\label{s2.2}

\begin{figure}[t]
\vskip 0.1 cm
\hskip -0.4 cm
\includegraphics[width=1.1\linewidth]{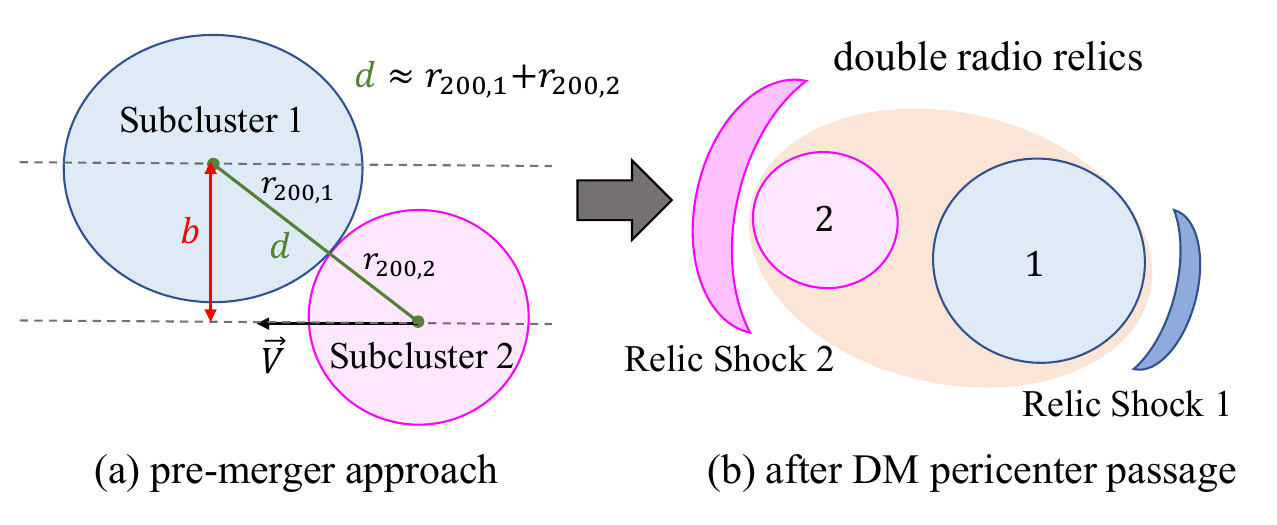}
\vskip -0.1 cm
\caption{Schematic picture of a binary merger with a nonzero impact parameter $b$ during (a) the pre-merger approach and (b) far after the pericenter passage. Here, $r_{200}$ is the radius within which the mean total density of the cluster attains 200 times the critical density, $\rho_c$, of the universe. The subscripts 1 and 2 denote heavier and lighter subclusters, respectively. A pair of merger shocks develop after the pericenter passage, and the one in front of subcluster 1 is labeled as relic shock 1 (blue), while the other in front of subcluster 2 is labeled as relic shock 2 (magenta).}\label{f1}
\end{figure}

\begin{deluxetable}{ccccccccc}\label{t1}
\tabletypesize{\footnotesize}
\tablecolumns{8}
\tablewidth{0pt}
\tablecaption{Merging Galaxy Cluster Sample}
\tablehead{
\colhead{No.$^a$}&
\colhead{$z_{\rm merg}^b$}&
\colhead{$z_{\rm relic}^c$}&
\colhead{{$\mathcal{M}_1^d$}}&
\colhead{{$\mathcal{M}_2^d$}}&
\colhead{$\mathcal{M}_1/\mathcal{M}_2$$^e$}&
\colhead{$b/r_{\rm vir,1}$$^f$}&
\colhead{$T_X$$^g$}}
\startdata
1 & 0.22 & 0.14 & 2.59 & 2.29 & 1.13 & 0.44 & 3.73 \\ 
2 & 0.60 & 0.50 & 0.94 & 0.65 & 1.46 & 0.93 & 2.48 \\ 
3 & 0.18 & 0.10 & 1.46 & 0.99 & 1.48 & 0.71 & 1.97 \\ 
4 & 0.50 & 0.42 & 2.09 & 1.32 & 1.58 & 0.55 & 3.21 \\ 
5 & 0.30 & 0.22 & 1.54 & 0.82 & 1.87 & 0.70 & 2.40 \\ 
6 & 0.26 & 0.18 & 1.51 & 0.68 & 2.20 & 0.46 & 2.75 \\ 
7 & 0.34 & 0.26 & 2.66 & 1.21 & 2.20 & 0.47 & 2.87 \\ 
8 & 0.26 & 0.18 & 1.51 & 0.61 & 2.49 & 0.42 & 3.04 \\ 
9 & 0.10 & 0.01 & 2.91 & 1.12 & 2.60 & 0.90 & 3.64 \\ 
10 & 0.10 & 0.01 & 3.63 & 1.23 & 2.96 & 0.30 & 3.14 \\ 
11 & 0.50 & 0.42 & 1.13 & 0.37 & 3.08 & 0.79 & 1.96 \\ 
12 & 0.60 & 0.50 & 1.46 & 0.39 & 3.75 & 0.45 & 2.55 \\ 
\enddata
\tablenotetext{a}{Clusters are arranged in ascending order of the mass ratio $\mathcal{M}_1/\mathcal{M}_2$.}
\tablenotetext{b}{Redshift of the pericenter passage, the epoch when the separation of the subcluster DM centers is shortest.}
\tablenotetext{c}{Redshift of merger shock identification and analysis, approximately the optimal time for merger shock surfaces to be observed as radio relics, found by viewing animations of simulated merging clusters over time. It corresponds to $\sim1$ Gyr after $z_{\rm merg}$.}
\tablenotetext{d}{Masses of two subclusters in units of $10^{14}\mathcal{M}_\odot$.}
\tablenotetext{e}{Mass ratio of two subclusters.}
\tablenotetext{f}{Impact parameter in units of the virial radius of the heavier subcluster, $r_{\rm vir,1}$.}
\tablenotetext{g}{X-ray-emissivity-weighted temperature of merged clusters at $z_{\rm relic}$ in units of keV.}
\vskip -1 cm
\end{deluxetable}

From seven different sets of simulations, we identify clusters with an X-ray-weighted temperature $T_X \gtrsim 2$~keV at redshift $z=0$. {Here, $T_X$ is calculated by weighting the gas temperature, $T$, with the bremsstrahlung emission, $\varepsilon_{\rm ff}\propto T^{1/2}\rho_{\rm gas}^2$, across the cluster volume (see \citet{kang1994} and \citet{Hong2015} for details), where $\rho_{\rm gas}$ is the gas density.} Among them, we find more than 20 clusters that have gone through major binary mergers. We construct a sample of 12 merging clusters satisfying the following three criteria: (1) mass ratio $1<\mathcal{M}_1/\mathcal{M}_2 \lesssim 4$; (2) impact parameter $b\lesssim r_{\rm vir,1}$; (3) $T_X \gtrsim 2$~KeV at the redshift of $z_{\rm relic}\lesssim0.5$. Hereafter, we denote heavier and lighter subclusters with the subscripts 1 and 2, respectively; $r_{\rm vir,1}$ is the virial radius of the heavier subcluster, and $z_{\rm relic}$ is the redshift at $\sim 1$~Gyr after the pericenter passage, the optimal time for merger shock surfaces to be observed as radio relics \citep[see][]{Ha2018}.

Figure \ref{f1} illustrates the schematic picture of a binary major merger, showing our definitions of the impact parameter and two relic shocks. In panel (a), which depicts a pre-merger phase, two subclusters are about to touch with a distance of $d\approx r_{200,1}+r_{200,2}$. Here, $r_{200}$ is the radius within which the mean total (gas plus DM) density of the cluster attains 200 times the critical density, $\rho_c$, of the universe. The impact parameter, $b$, is defined as the distance perpendicular to the paths of the subcluster centers. Typically, this corresponds to $\sim1.5-2$ Gyr before the pericenter passage. Panel (b) depicts a stage following the pericenter passage, showcasing two merger shocks, relic shock 1 in front of subcluster 1 and relic shock 2 in front of subcluster 2. We use the term ``relic shock'' because, as noted in the introduction, merger shocks are often observed as radio relics. {Given that most binary radio relics in major mergers are observed in the post-pericenter passage phase \citep{Golovich2019}, our analysis focuses on the stage shown in panel (b).}

Table \ref{t1} lists the 12 merging clusters in our sample. Here, $r_{\rm vir} \approx 1.36\ r_{200}$ \citep[e.g.,] []{Roncarelli2006, Reiprich2013}, and $\mathcal{M}_1$ and $\mathcal{M}_2$ are the total (gas plus DM) mass contained within the subcluster's virial radius at the epoch of $d\approx r_{200,1}+r_{200,2}$. {The masses and temperatures of the merging clusters in this study are somewhat lower than typical values for observed merging clusters (see Section \ref{s4}) due to the limited computational volume in our simulations. Therefore, we present results primarily for normalized dimensionless quantities, assuming these are approximately scale-independent.}

\begin{figure}[t]
\vskip 0.1 cm
\hskip -0.3 cm
\includegraphics[width=1.1\linewidth]{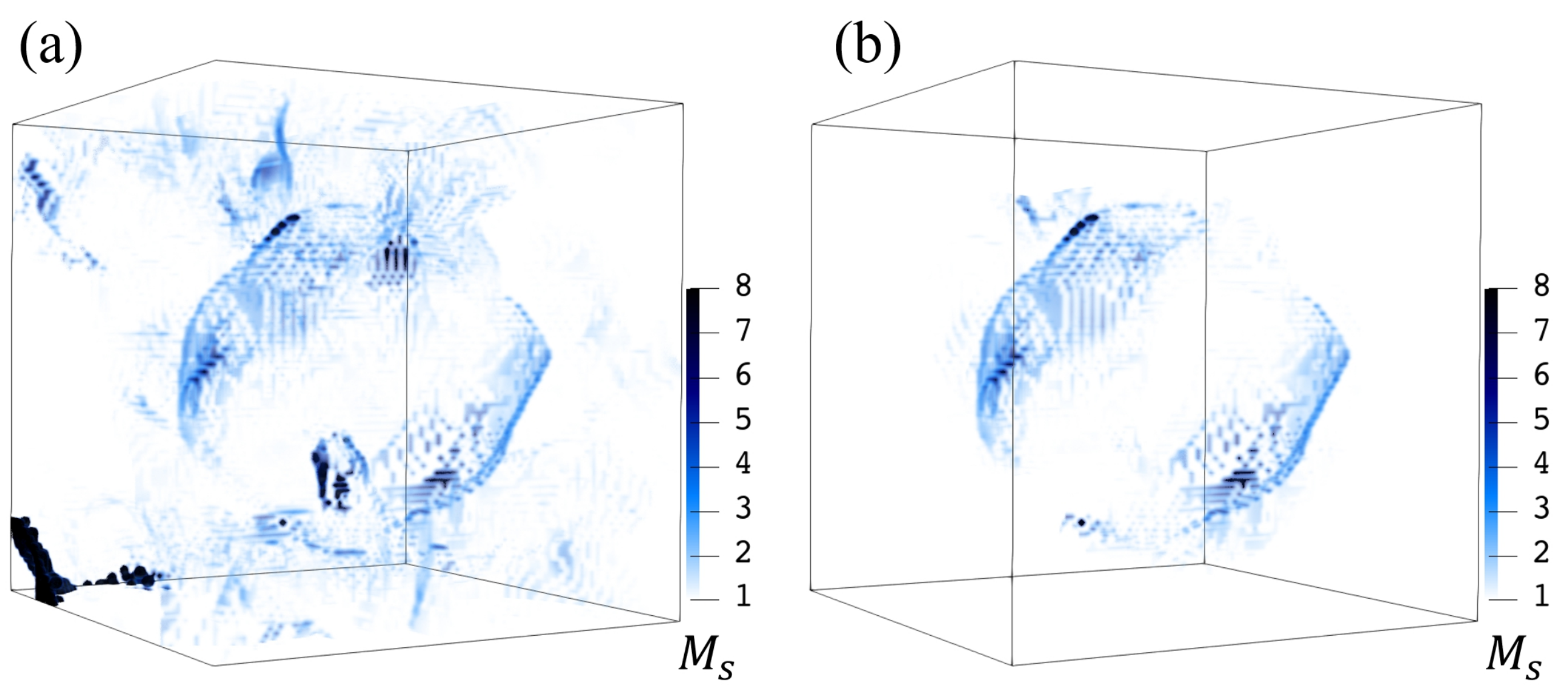}
\vskip -0.1 cm
\caption{3D view of shock zones in the comoving volume of $(4.88\ {\rm Mpc})^3$ that encompasses Cluster 3 (see Table \ref{t1}). Color displays the Mach number of shock zones. (a) All shock zones within the volume. (b) Shock zones  satisfying the density/temperature and distance criteria (see the main text). Here, only the shock zones within $\left<d_s\right>\pm1\sigma_{d_s}$ are included, where $\left<d_s\right>$ is the mean distance of shock zones from the X-ray center of the cluster, and $\sigma_{d_s}$ is its standard deviation. The images are captured at $z_{\rm relic}$.}\label{f2}
\end{figure}

\begin{figure*}[t] 
\vskip 0.1 cm
\hskip 0.2 cm
\includegraphics[width=1\linewidth]{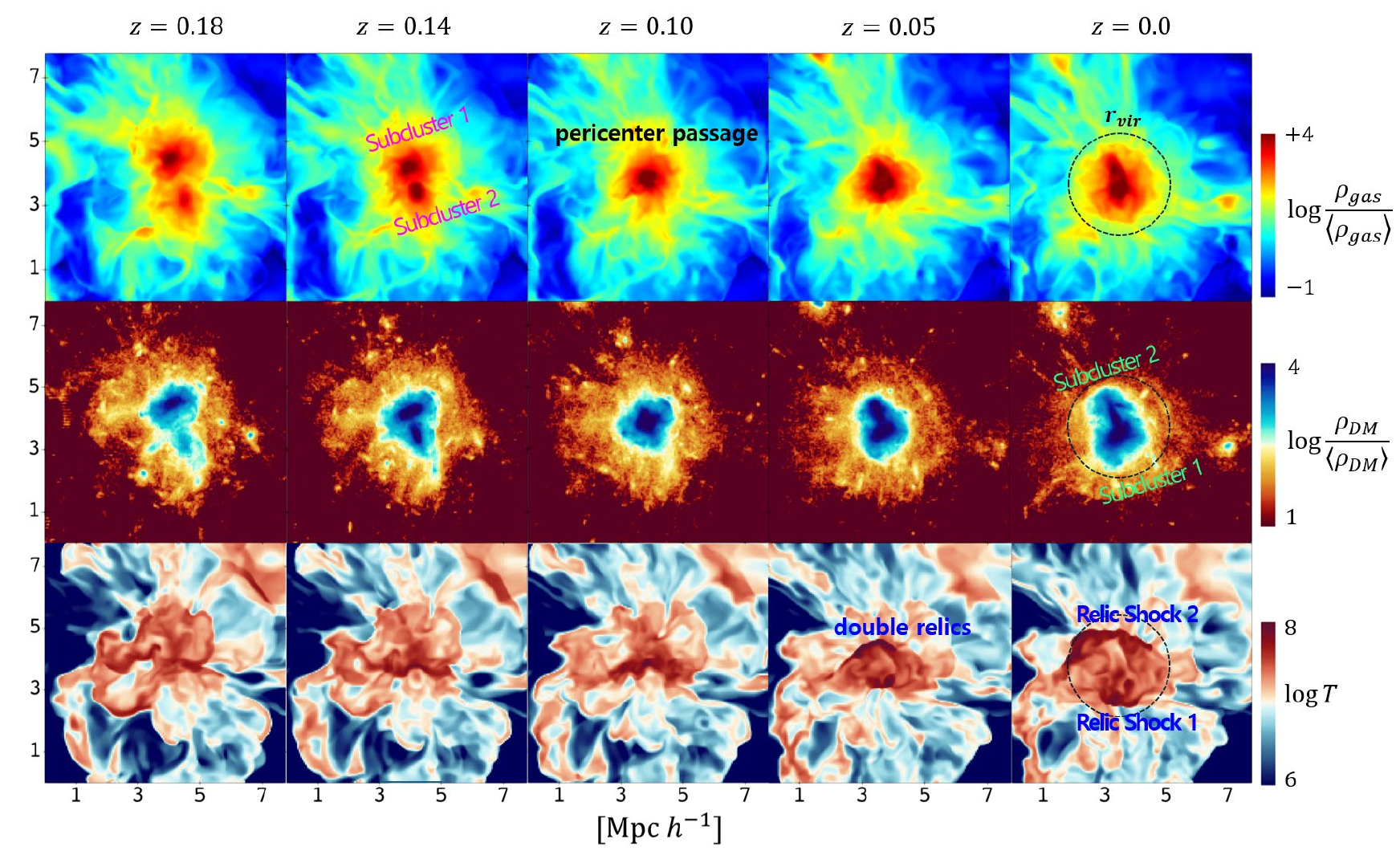}
\vskip -0.2 cm
\caption{Merging process in Cluster 10 with $\mathcal{M}_1/\mathcal{M}_2\simeq2.96$ and $b/r_{\rm vir,1}\simeq0.3$ (see Table \ref{t1}). From top to bottom, 2D slice images of the gas density, $ \rho_{\rm gas}/\left<\rho_{\rm gas}\right>$, the DM density, $\rho_{\rm DM}/\left<\rho_{\rm DM}\right>$, and the gas temperature, $T$, in the comoving area of $(8~h^{-1}{\rm Mpc})^2$ around the cluster are shown. Here, $\left<\rho_{\rm gas}\right>$ and $\left<\rho_{\rm DM}\right>$ are the mean gas and DM densities of the universe, and $T$ is in units of Kelvin. The pericenter passage occurs at $z\approx0.1$, while a pair of merger shocks would appear as radio relics at $z\lesssim 0.05$. Relic shock 2 in front of subcluster 2 is larger than relic shock 1 in front of subcluster 1. The black dot at $z=0$ draws the circle of the virial radius, $r_{\rm vir}$.}\label{f3}
\end{figure*}

\subsection{Shock Zones and Merger Shock Surfaces}\label{s2.3}

In merged clusters, we identify shocks using the algorithm introduced in \citet{ryu2003}. Along the axis directions, grid zones are tagged as shock zones if they satisfy the following three conditions: (1) $\nabla\cdot{v} < 0$, i.e., locally converging flow; (2) $\Delta T\times\Delta\rho > 0$, i.e., the same sign of gas temperature and density gradients; and (3) $\left\vert \Delta \log{T} \right\vert > 0.11 $, i.e., the temperature jump larger than that of Mach number $M_s=1.3$. Then, $M_s$ is calculated by solving the Rankine-Hugoniot temperature jump, $T_2/T_1 = (5M_s^2-1)(M_s^2+3)/(16M_s^2)$. Here, the subscripts 1 and 2 represent the preshock and postshock quantities, respectively. After applying these procedures for all three axis directions, the Mach number of a shock zone is defined as the maximum value of the three Mach numbers along the $x$, $y$ and $z$ axes, $M_s = \max(M_{s,x},M_{s,y},M_{s,z})$. Many grid zones are identified to contain weak shocks, but weak shocks are not energetically important. Thus, we only consider the shock zones with $M_s \ge 1.5$.

In the ICM, shocks can be induced by dynamical activities other than mergers of subclusters, such as turbulent flow motions and the infall of the WHIM along filaments. To isolate shock zones that belong to the merger-driven shock surfaces, we employ the following steps. In the first step, considering merger shocks typically appear around the virial radius (see the next section), we select shock zones satisfying (1) $-1.4 \le \log_{10} \rho_1/\langle \rho \rangle \le 0.3$ and (2) $-1.0 \le \log_{10} T_1/ \langle T \rangle \le 0.1$. Here, $\rho_1$ and $T_1$ represent the preshock density and temperature, respectively, and $\langle \rho \rangle$ and $\langle T \rangle$ are the mean values calculated within the virial radius of each cluster. The upper bounds for $\rho_1$ and $T_1$ help eliminate turbulence-induced shocks in the dense and hot ICM inside the virial radius, while the lower bounds are imposed to exclude shocks induced by the infall of the WHIM along filaments in cluster outskirts. These criteria have been empirically determined by examining the merger shocks at $z_{\rm relic}$. In the second step, using the selected shock zones, we estimate the mean distance $\langle d_s \rangle $ of each relic shock surface from the cluster center and its standard deviation, $\sigma_{d_s}$. Throughout the paper, the cluster center is defined as the peak of X-ray emission within the merged cluster. Only those within $\left<d_s\right>\pm1\sigma_{d_s}$ are chosen as parts of the relic shock surface. We then iteratively update $\langle d_s \rangle $ and $\sigma_{d_s}$ and refine the list of constituent shock zones to achieve the best result. 

Figure \ref{f2} illustrates three-dimensional (3D) views of the shock surfaces in Cluster 3 at $z_{\rm relic}\approx 0.1$ (see Table \ref{t1}). The left panel displays all shock zones within the volume, whereas the right panel shows the shock zones after the refinement according to the above criteria. Below we examine the properties of refined merger shock surfaces.

\section{Results}\label{s3}

\subsection{Morphological Properties of Merger Shock Surfaces}\label{s3.1}

\begin{figure*}[t]
\vskip 0.1 cm
\hskip 0 cm
\includegraphics[width=1.0\linewidth]{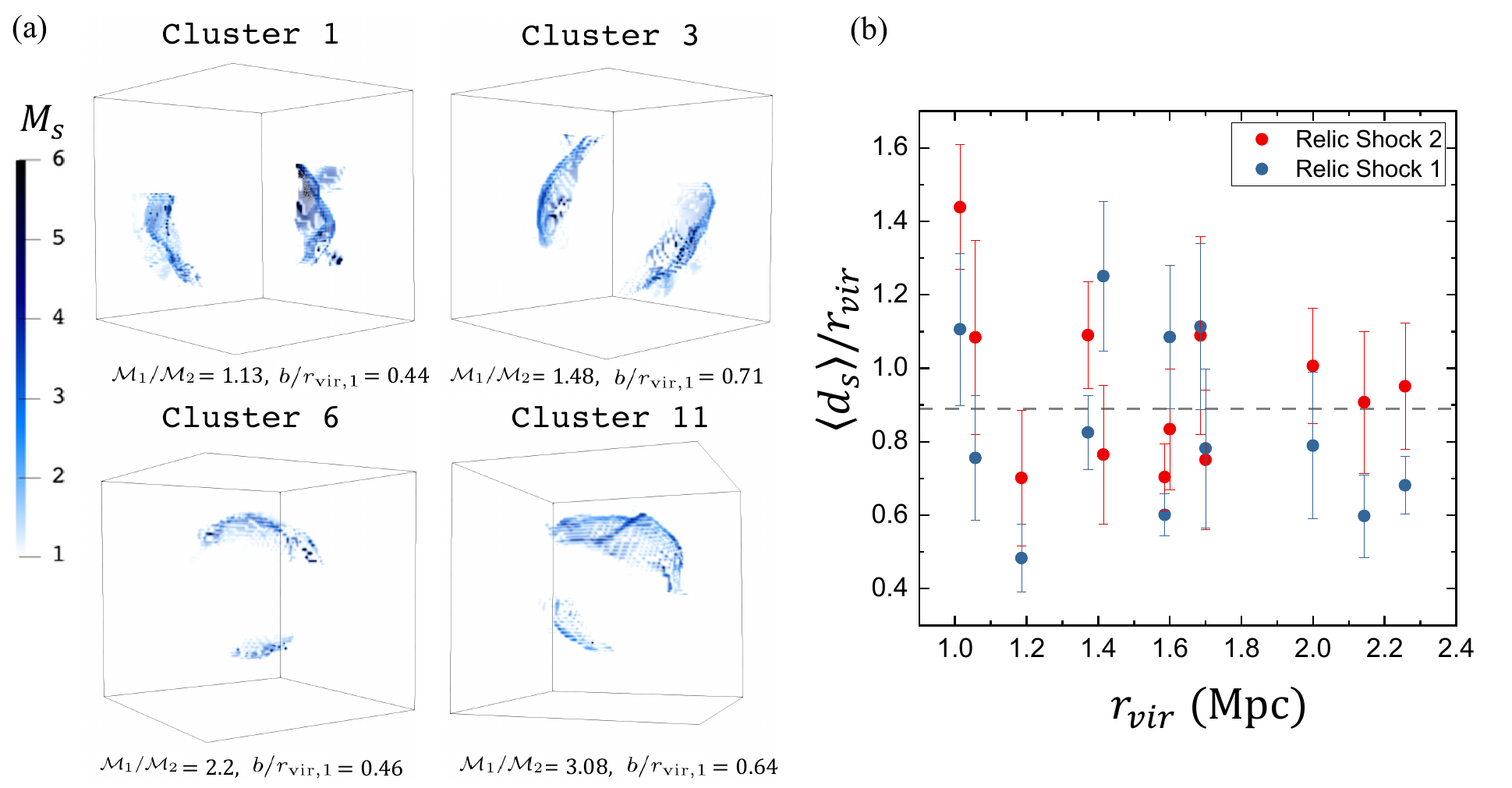}
\vskip -0.1 cm
\caption{(a) 3D view of merger shocks at $z_{\rm relic}$ for four representative clusters. Color displays the Mach number of shock zones. The comoving volume of the boxes ranges approximately $(4-5~{\rm Mpc})^3$; the box size varies to optimize the visualization of shock surfaces. Merger shocks for all 12 sample clusters in Table \ref{t1} are provided in Figure \ref{fA1}. (b) Mean distance, $\langle d_{s} \rangle$, between the merger shock surface and the X-ray center of the merged cluster, normalized to the virial radius of each cluster, $r_{\rm vir}$, versus $r_{\rm vir}$ for 24 merger shocks in Figure \ref{fA1}. Here, $r_{\rm vir}$ is given as the proper distance. The vertical error bars represent the $\pm1\ \sigma_{d_s}$ deviation. The horizontal dashed line indicates the average value of $\langle d_{s} \rangle / r_{\rm vir}\approx 0.9$. Blue dots and lines represent relic shock 1, while red dots and lines represent relic shock 2, the same colors as in Figure \ref{f1}. The same color scheme is used in subsequent figures.}\label{f4}
\end{figure*}

Mergers of subclusters take place during the hierarchical formation of the LSS of the universe; the resulting merger shocks, which develop in the inherently turbulent ICM, usually have complex morphology. Figure \ref{f3} displays two-dimensional (2D) slices for the gas density (top panels), DM density (middle panels), and gas temperature (bottom panels) in Cluster 10 (see Table \ref{t1}). At $z=0.18$, two prominent clumps approach each other with $\mathcal{M}_1/\mathcal{M}_2\simeq2.96$ and $b/r_{\rm vir,1}\simeq0.3$. As these clumps close in with a relative velocity of about $10^3~{\rm km~s^{-1}}$, shocks develop along the merger axis and become visible at $z=0.14$. The epoch near the pericenter passage occurs around $z=0.1$. Following this, two merger shocks are observed propagating outward over $z=0.1-0.0$. The shock surfaces and the turbulent nature of the ICM are most clearly depicted in the temperature images. On the other hand, the gas density images reveal multiple minor mergers and secondary infalls that contribute to turbulent flow motions in the ICM. {Although not shown here, the shock speed, $v_s$, and also $M_s$ initially increase with time and then tend to converge, though significant fluctuations are observed \citep[see Figure 5 of][]{Ha2018}.}

Figure \ref{f4}(a) presents 3D views of merger shocks at $z_{\rm relic}$ for four of our sample clusters (Clusters 1, 3, 6, and 11), highlighting pairs of shock surfaces moving in opposite directions. Merger shocks for all 12 sample clusters are shown in Figure \ref{fA1}. In Clusters 6 and 11 with relatively large $\mathcal{M}_1/\mathcal{M}_2$, a larger shock is in front of the lighter dark matter clump (relic shock 2), while a smaller shock appears in front of the heavier dark matter clump (relic shock 1). Additionally, the figure illustrates the asymmetry, caused by nonzero impact parameters and the turbulent ICM.

Figure \ref{f4}(b) plots the mean distance of merger shocks from the cluster center, $\langle d_{s} \rangle$, as a function of the virial radius of the merged cluster, $r_{\rm vir}$, at $z_{\rm relic}$. For our sample clusters with $r_{\rm vir} \approx 1.0-2.3~{\rm Mpc}$ (proper distance), $\langle d_{s} \rangle/r_{\rm vir}$ ranges $\sim0.6-1.2$ with an average value of $\sim0.9$. This indicates that at $z_{\rm relic}$, $\sim 1$ Gyr after the pericenter passage, although there is a significant scatter, merger shocks are typically found around the virial radius, consistent with observations of radio relic locations (see the introduction). {It is worth noting that \citet{Zhang2019} proposed a “habitable zone” at distances $\gtrsim r_{\rm 500}$, or $\gtrsim 0.5\ r_{\rm vir}$, beyond which moderately strong shocks appear. Here, $r_{\rm 500}$ is the radius within which the mean total density is 500 times $\rho_c$. Although their simulation setup and merger parameters differ from ours, the merger shocks in Figure \ref{f4}(b) are indeed in the habitable zone, aligning with their suggestion.} Additionally, the figure shows that there is almost no or only weak dependence of $\langle d_{s}\rangle/r_{\rm vir}$ on $r_{\rm vir}$, implying that on average, $\langle d_{s} \rangle$ tends to be larger in more massive clusters with larger $r_{\rm vir}$. 

\begin{figure*}[t]
\vskip 0.1 cm
\hskip -0.1 cm
\includegraphics[width=1.0\linewidth]{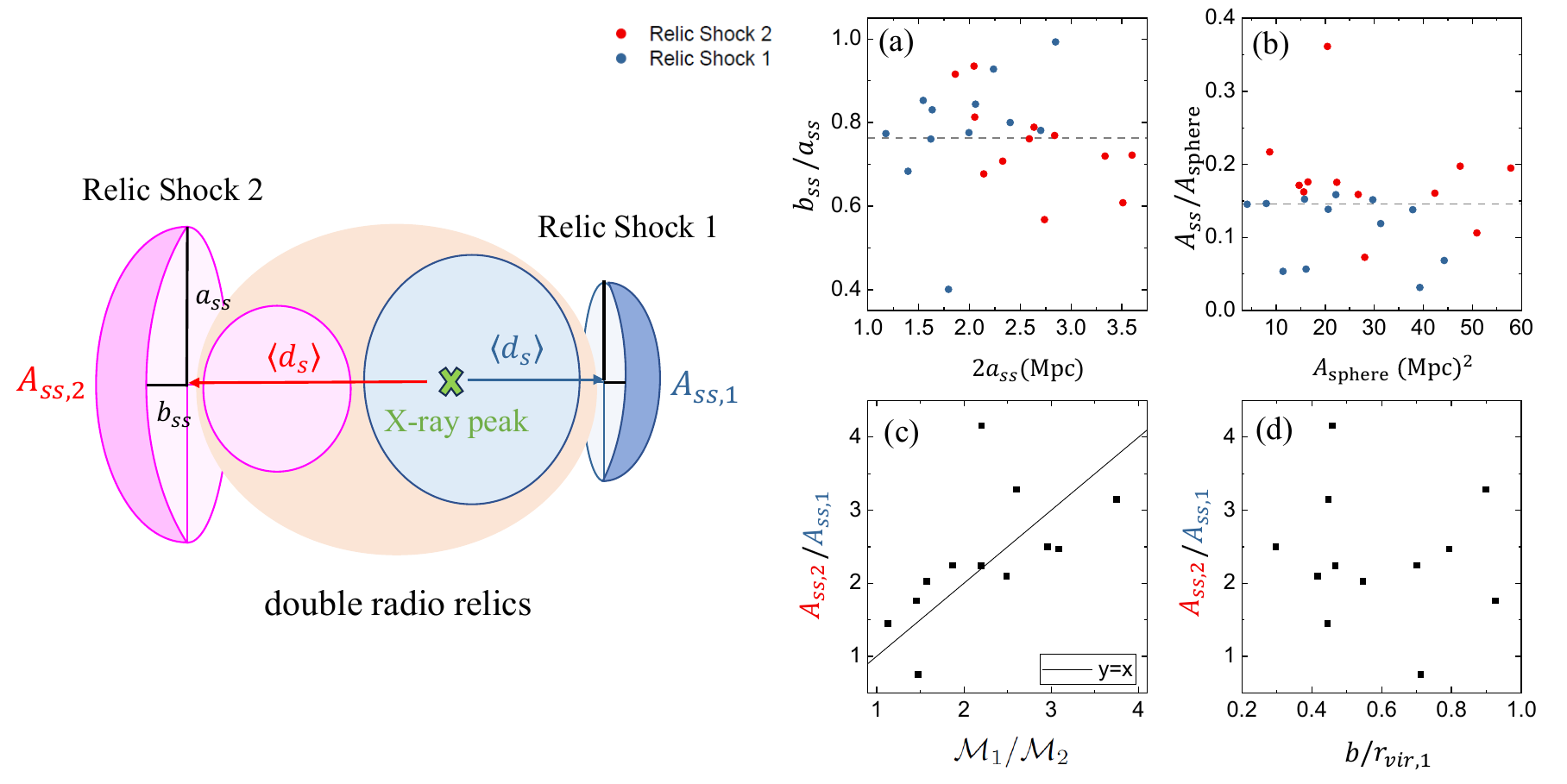}
\vskip -0.2 cm
\caption{Left Panel: Schematic picture showing the geometry of relic shocks in a binary merger far after the pericenter passage. The shock surfaces are approximated as ellipses with major axis $a_{\rm ss}$ and minor axis $b_{\rm ss}$. $A_{\rm ss}$ denotes the surface area of shocks and is calculated as $A_{\rm ss}\approx 1.19(\Delta l)^2N_{\rm sh}$, where $N_{\rm sh}$ is the number of shock zones associated with a particular shock surface. Right Panel: (a) Axial ratio of merger shock surfaces, $b_{\rm ss}/a_{\rm ss}$, for the 24 merger shocks in Figure \ref{fA1}. (b) Ratio of the area of merger shock surfaces, $A_{\rm ss}$, to the area of the shock spheres of radius $\langle d_{s}\rangle$, $A_{\rm sphere}=4\pi\langle d_{s}\rangle^2$. (c)-(d) Ratio of $A_{\rm ss,2}/A_{\rm ss,1}$ versus the mass ratio of two subclusters, $\mathcal{M}_1/\mathcal{M}_2$, and the normalized impact parameter, $b/r_{\rm vir,1}$, for 12 sample clusters. The horizontal dashed lines in (a)-(b) denote the average values for the 24 merger shocks. The solid line in panel (c) draws $y=x$.}\label{f5}
\end{figure*}

\begin{figure*}[t]
\vskip 0.2 cm
\hskip 0 cm
\includegraphics[width=1.0\linewidth]{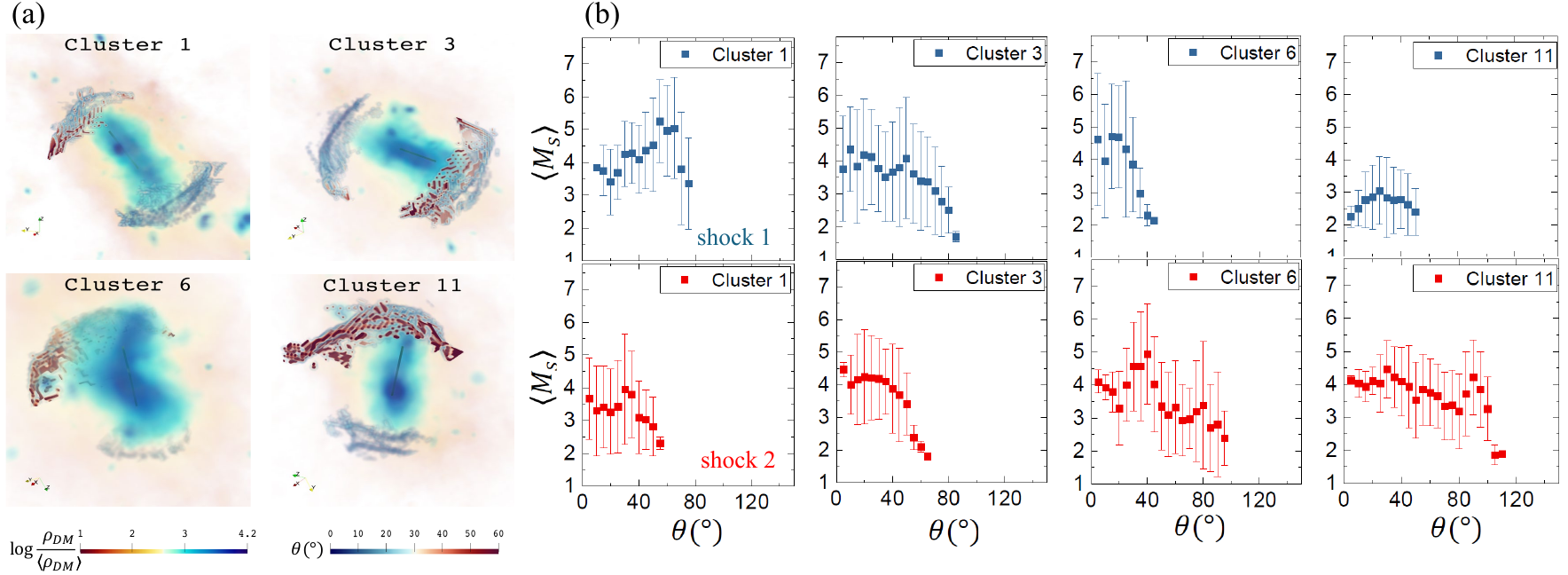}
\vskip -0.0 cm
\caption{(a) 3D view of merger shocks, superimposed on 3D volume-rendered images of the DM density, $\rho_{\rm DM}/\langle\rho_{\rm DM}\rangle$, for the four clusters shown in Figure \ref{f4}. The color of merger shocks displays the ``position angle'' of shock surfaces, $\theta$, which is the angle between the axis passing through the two DM density peaks (indicated by black line segments) and the line connecting the midpoint of these peaks to shock zones. The images are at $z_{\rm relic}$ as in Figure \ref{f4}, but with different viewing angles. The corresponding images for all 12 sample clusters are provided in Figure \ref{fA2}. (b) Average Mach number of shock zones associated with relic shocks 1 (blue) and 2 (red) as a function of $\theta$ with bins of $\Delta \theta = 5^{\circ}$ for the four sample clusters. Error bars represent $\pm 1\ \sigma$ deviations of $M_s$. Similar plots for all 12 sample clusters are shown in Figure \ref{fA3}.}\label{f6}
\end{figure*}

The left panel of Figure \ref{f5} schematically depicts the geometry of merger shock surfaces as elliptical sections of spherical shells at distance $\langle d_s \rangle$, described by their semi-major axis $a_{\rm ss}$ and semi-minor axis $b_{\rm ss}$. The major axis ranges $2a_{\rm ss} \approx 1.2-3.6~{\rm Mpc}$, while the minor axis ranges $2b_{\rm ss} \approx0.7-2.8~{\rm Mpc}$, in terms of the proper distance. As shown in Figure \ref{f5}(a), the aspect ratio is generally $b_{\rm ss}/a_{\rm ss}\gtrsim0.6$ with an average of $\sim0.77$, except for an outlier with $b_{\rm ss}/a_{\rm ss}\approx0.4$ observed for relic shock 1 in Cluster 5. So, while most merger shock surfaces normally appear moderately elongated, highly elongated, stripe-like surfaces can also form depending on the merger history of clusters as well as on the flow dynamics in the turbulent ICM.

We define the ``shock sphere'' as a sphere with radius $\langle d_s \rangle$ and surface area $A_{\rm sphere} = 4\pi \langle d_s \rangle^2$. For merger shock surfaces, the area is calculated as $A_{\rm ss} \approx 1.19(\Delta l)^2 N_{\rm sh}$, where $N_{\rm sh}$ is the number of shock zones associated with a given merger shock. The factor of 1.19 accounts for the mean projected area within a 3D zone with random shock normal orientations. In Figure \ref{f5}(b), the normalized area of merger shock surfaces, $A_{\rm ss} / A_{\rm sphere}$, is plotted. For most relic shocks, $A_{\rm ss} / A_{\rm sphere}$ ranges $\sim0.05 - 0.2$, with the exception of relic shock 2 in Cluster 8, which has $A_{\rm ss} / A_{\rm sphere} \approx 0.36$. The average value is $A_{\rm ss} / A_{\rm sphere}\sim0.15$, indicating that merger shock surfaces typically occupy a small fraction of the surface area of shock spheres.

The morphological properties of merger shock surfaces can vary depending on merger parameters. In idealized binary mergers with subclusters in hydrostatic equilibrium, it was shown that the mass ratio influences the relative size of two merger shock surfaces, while the impact parameter affects the degree of asymmetry around the merger axis \citep[e.g.,][]{vanWeeren2011a,Molnar2017}. Figure \ref{f5}(c) and (d) show the ratio of the surface areas of two merger shocks, $A_{\rm ss,2}/A_{\rm ss,1}$, as a function of merger parameters, $\mathcal{M}_1/\mathcal{M}_2$ and $b/r_{\rm vir,1}$, for our sample clusters. The ratio $A_{\rm ss,2}/A_{\rm ss,1}$ tends to increase with increasing $\mathcal{M}_1/\mathcal{M}_2$, consistent with the findings of previous studies on idealized binary mergers mentioned above. In contrast, while there is considerable scatter, no clear dependence of $A_{\rm ss,2}/A_{\rm ss,1}$ on $b/r_{\rm vir,1}$ is found.

\subsection{Mach Numbers as a Function of Angular Distance from the Merger Axis}\label{s3.2}

As an effort to describe the morphological characteristics of relic shocks in the turbulent ICM, we examine the distribution of $M_s$ in merger shock surfaces as a function of the position angle, $\theta$, which is defined as the angle between the axis connecting two DM density peaks and the line extending from the midpoint of these peaks to the given shock zone. Figure \ref{f6}(a) depicts the 3D distribution of $\theta$ for merger shocks, on top of the DM density distribution, for the four clusters shown in Figure \ref{f4}(a). The corresponding images for all 12 sample clusters are available in Figure \ref{fA2}. Figure \ref{f6}(b) plot the average and standard deviation of $M_s$ in bins of $\Delta\theta$ as a function of $\theta$ for shock zones that belong to relic shocks 1 (blue) and 2 (red) in the four sample clusters. Similar plots for all 12 sample clusters are provided in Figure \ref{fA3}.

In the idealized scenario of a spherical bow shock propagating through a uniform medium, the Mach number distribution is expected to vary as $\langle M_s \rangle \propto \cos{\theta}$ {\citep[see, e.g.,][for further discussion]{Petrinec1997}}. {\citet{Russell2022} investigated the angular dependence of the Mach number for a merger shock in Abell 2146 using X-ray data, finding that it roughly follows the cosine behavior. In contrast, our study examines the angular dependence of the Mach number in 3D.} As shown in Figures \ref{f6}(b) and \ref{fA3}, while some merger shocks, such as relic shock 2 of Cluster 3, exhibit behaviors close to the idealized pattern, many do not follow this trend. Specifically, some shocks exhibit peaks at large nonzero $\theta$ values, rather than around $\theta \sim 0^{\circ}$, with a roughly symmetrical decline on either side of the peak (e.g., relic shock 1 of Cluster 11). Others display significant fluctuations in the Mach number distribution with no distinct peak. As a matter of fact, in most cases, the axis connecting DM peaks does not go through the center of shock surfaces. {This is partly due to nonzero impact parameters in our sample clusters and also due to the fact that merger shock surfaces are disturbed by turbulent flow motions and infalling clumps in the ICM as they propagate from the core to outskirts. Consequently, the center of the shock surface fluctuates and becomes misaligned with the axis connecting the DM peaks.} For example, for relic shock 1 of Cluster 1, the shock surface is significantly tilted, so there is no shock zone for $\theta\lesssim5^{\circ}$. We point out that Cluster 1 has experienced a major merger at $z = 0.22$, followed by a minor merger with a mass ratio of $\sim 6$ in the propagation direction of relic shock 1 at $z = 0.18$. And the shock properties are analyzed at $z_{\rm relic} = 0.14$. In this relic shock, the shock surface morphology is affected by the late minor merger (see the shock on the right hand side of Cluster 1 in Figure \ref{f4}(a)) and offset from the axis connecting DM peaks.

\subsection{Mach Number Distribution on Merger Shock Surfaces}\label{s3.3}

\begin{figure*}[t]
\vskip 0.1 cm
\hskip -0.0 cm
\includegraphics[width=1.0\linewidth]{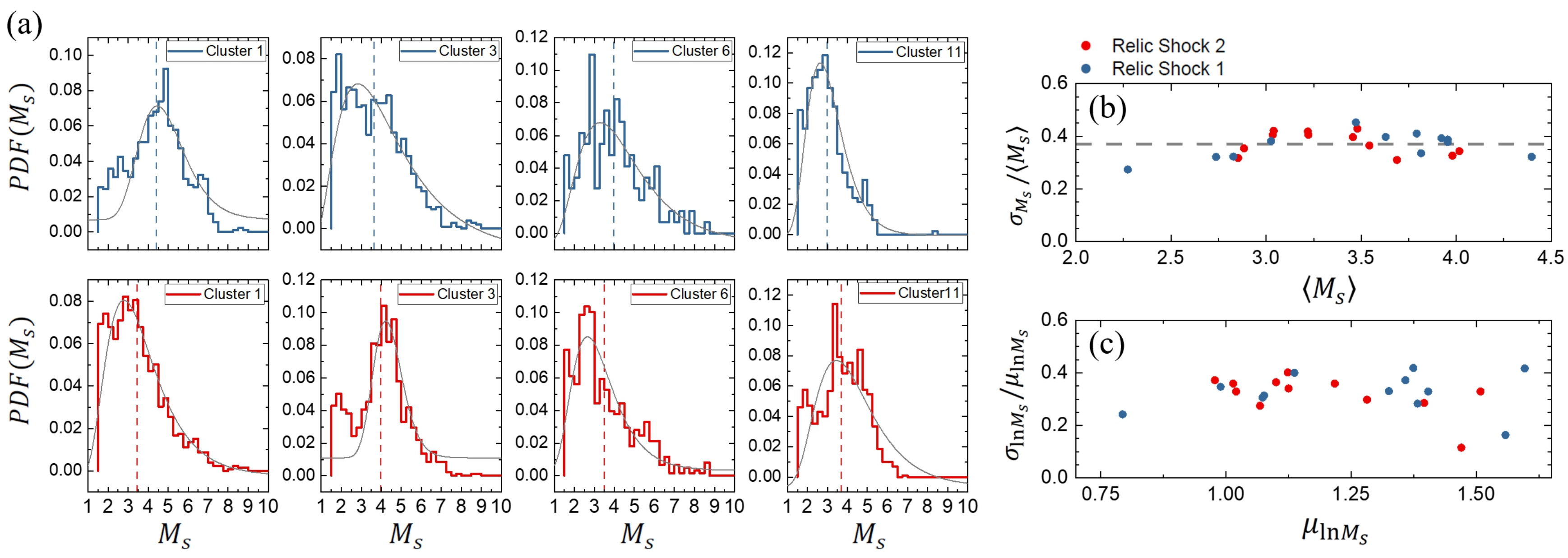}
\vskip -0.2 cm
\caption{(a) PDF of the Mach number of shock zones associated with relic shocks 1 (blue) and 2 (red) in the four clusters shown in Figure \ref{f4}. The vertical dashed lines denote the shock-surface-area-weighted average Mach number, $\langle M_{s} \rangle_{\rm area}$, while the gray solid lines show the log-normal fitting. The corresponding PDFs for all 12 sample clusters are provided in Figure \ref{fA4}. (b) Standard deviation normalized to the mean value, $\sigma_{M_s}/\langle M_{s} \rangle_{\rm area}$, as a function of $\langle M_{s} \rangle_{\rm area}$ for the PDFs of 24 merger shocks in 12 sample clusters. The subscript `area' is omitted for simplicity. The horizontal dashed line represents the average value of $\sigma_{M_{s}}/\langle M_{s} \rangle \approx 0.37$. (c) $\sigma_{\ln M_s}/\mu_{\ln{M_{s}}}$ versus $\mu_{\ln{M_{s}}}$, where $\mu_{\ln{M_{s}}}$ and $\sigma_{\ln M_s}$ are the mean and standard deviation for the fitted log-normal distribution of the Mach number.}\label{f7}
\end{figure*}

The PDFs for $M_s$ of shock zones across merger shock surfaces are shown in Figure \ref{f7}(a) for the four representative clusters. The corresponding PDFs for all 12 sample clusters are provided in Figure \ref{fA4}. In general, the PDFs exhibit positively skewed distributions. We fit the PDFs to a log-normal function, with the fittings shown as gray lines in the figures. The log-normal fits peak around $M_{s,\rm peak} \approx 2-4.5$, with a high Mach number tail extending up to $M_s \sim 10$. This tail is primarily due to intermittent accretions of the WHIM, as merger shocks are often attached to infalling filaments. These Mach number distributions agree with the findings of previous numerical studies \citep[e.g.,][]{Botteon2020,Wittor2021}.

For the further analysis of their statistical characteristics, we calculate (1) the mean and standard deviation of $M_s$, $\langle M_{s} \rangle_{\rm area}$ and $\sigma_{M_{s}}$, using the actual PDFs (blue and red lines in Figures \ref{f7}(a) and \ref{fA4}, and (2) the mean and standard deviation, $\mu_{\ln{M_{s}}}$ and $\sigma_{\ln{M_{s}}}$, using the fitted log-normal distributions (gray lines). The subscript `area' is added to $\langle M_{s} \rangle_{\rm area}$ to specify that it is actually the shock-surface-area-weighted average Mach number. Figures \ref{f7}(b) and (c) demonstrate that $\sigma_{M_s} / \langle M_{s} \rangle_{\rm area}$ and $\sigma_{\ln{M_{s}}}/\mu_{\ln{M_{s}}}$ are largely independent of $\langle M_{s} \rangle_{\rm area}$ and $\mu_{\ln{M_{s}}}$, with average values of $\sigma_{M_{s}}/\langle M_{s} \rangle_{\rm area} \approx 0.37$ and $\sigma_{\ln{M_{s}}}/\mu_{\ln{M_{s}}} \approx 0.33$, respectively. This implies that the PDFs of $M_s$ are more or less universal, although there is scatter owing to diverse merger histories of clusters, including the infall of small clumps, as well as turbulent flow motions in the ICM.

\begin{figure}[t]
\vskip 0cm
\hskip 0 cm
\includegraphics[width=0.9\linewidth]{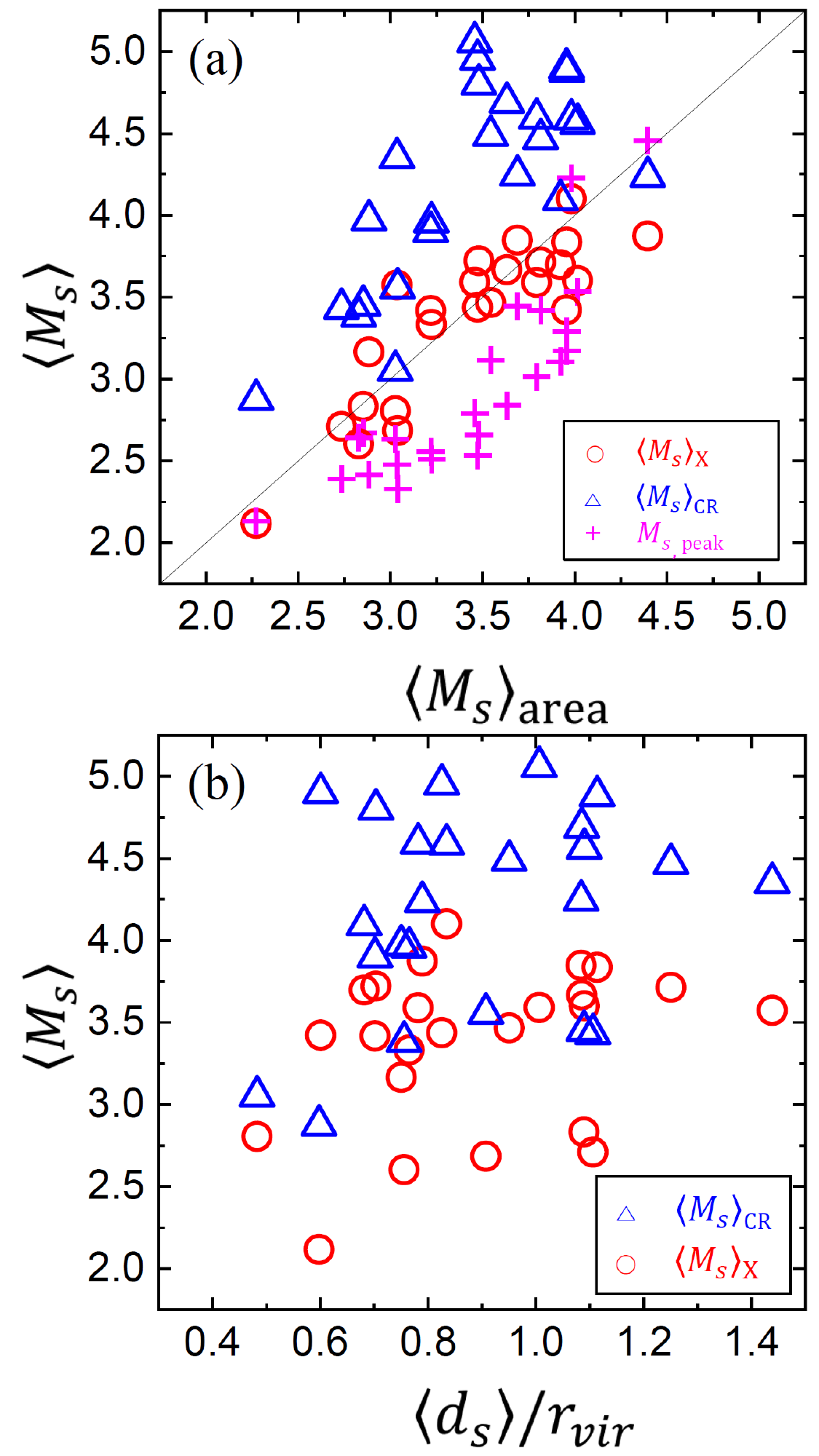}
\vskip -0.2 cm
\caption{(a) Peak of the Mach number PDF, $M_{s,{\rm peak}}$ (magenta), X-ray-emissivity-weighted average Mach number, ${\langle M_s \rangle}_X$ (red), and CR-energy-flux-weighted average Mach number, ${\langle M_s \rangle}_{\rm CR}$ (blue), versus shock-surface-area-weighted average Mach number, ${\langle M_s \rangle}_{\rm area}$, for 24 merger shocks. (b) ${\langle M_s \rangle}_X$ (red) and ${\langle M_s \rangle}_{\rm CR}$ (blue) as a function of the normalized distance of merger shocks from the cluster center, $\left<d_s\right>/r_{\rm vir}$.}\label{f8}
\end{figure}

\subsection{Average Mach Numbers of Merger Shock Surfaces} \label{s3.4}

In addition to $M_{s,\rm peak}$ and $\langle M_{s} \rangle_{\rm area}$, the Mach number distribution of shock zones on merge shock surfaces can be characterized by other representative values, such as the average Mach numbers weighted with observable quantities. Specifically, we examine the X-ray-emissivity-weighted average Mach number, ${\langle M_s \rangle}_X$, and the CR-energy-flux-weighted average Mach number, ${\langle M_s \rangle}_{\rm CR}$. Here, ${\langle M_s \rangle}_{\rm X}$ is obtained by weighing $M_s$ with the bremsstrahlung emission, $\varepsilon_{\rm ff}$, over shock zones on the given shock surface. For ${\langle M_s \rangle}_{\rm CR}$, $M_s$ is weighted by the CR energy flux, $f_{\rm CR} = \eta_e(M_s) \cdot f_{\phi}$, where $f_{\phi} = (1/2)\rho_1 {v_s}^3$ represents the shock kinetic energy flux. The CR electron acceleration efficiency is approximated as $\eta_e\approx 0.01 \eta$, where the CR proton acceleration efficiency ranges $\eta(M_s) \approx 3.6 \times 10^{-3} - 0.01$ for $2.25 \lesssim M_s \lesssim 5.0$ \citep{Ryu2019}. In this study, we regard ${\langle M_s \rangle}_{\rm CR}$ as a proxy for radio-emissivity-weighted average Mach number, ${\langle M_s \rangle}_{\rm radio}$.

In general, $M_{s,\rm peak}$ tends to be slightly smaller than $\langle M_{s} \rangle_{\rm area}\approx\ 2.3-4.4$, as expected with the log-normal shape of the Mach number PDFs, as shown in Figure \ref{f8}(a). The figure also illustrates that ${\langle M_s \rangle}_{\rm X} \sim 2-4$ is comparable to ${\langle M_s \rangle}_{\rm area}$. On the other hand, ${\langle M_s \rangle}_{\rm CR} \sim 3-5$ is greater than both ${\langle M_s \rangle}_{\rm area}$ and ${\langle M_s \rangle}_{\rm X}$. This trend reflects our CR acceleration model, $\eta(M_s)$, which predicts higher efficiencies for higher $M_s$, as noted in the introduction. {These results are consistent with those of previous studies. For example, using simulations with a cosmological magnetohydrodynamic code, ENZO, \citet{Wittor2021} reported average Mach numbers of ${\langle M_s \rangle}_{\rm X-ray} \approx 1-4$ and ${\langle M_s \rangle}_{\rm radio} \approx 3-5$ for merger shocks; in their work, ${\langle M_s \rangle}_{\rm X-ray}$ and ${\langle M_s \rangle}_{\rm radio}$ were estimated mimicking observations, including projection effects and synchrotron modeling.} In summary, $M_{s,\rm peak} \lesssim {\langle M_s \rangle}_{\rm area} \approx {\langle M_s \rangle}_{\rm X} \lesssim {\langle M_s \rangle}_{\rm CR}$, indicating that while ${\langle M_s \rangle}_{\rm X}$ provides a reasonable estimate of ${\langle M_s \rangle}_{\rm area}$, ${\langle M_s \rangle}_{\rm CR}$ is typically higher. However, there are exceptions; in relic shock 1 of Cluster 1, where a secondary minor merger has disturbed the shock surface, ${\langle M_s \rangle}_{\rm CR}$ is close to ${\langle M_s \rangle}_{\rm area}$, while ${\langle M_s \rangle}_{\rm X}$ is smaller (see the points with ${\langle M_s \rangle}_{\rm area}=4.4$ in the figure). 

\begin{figure}[t]
\vskip 0.2cm
\hskip -0.6 cm
\includegraphics[width=1.08\linewidth]{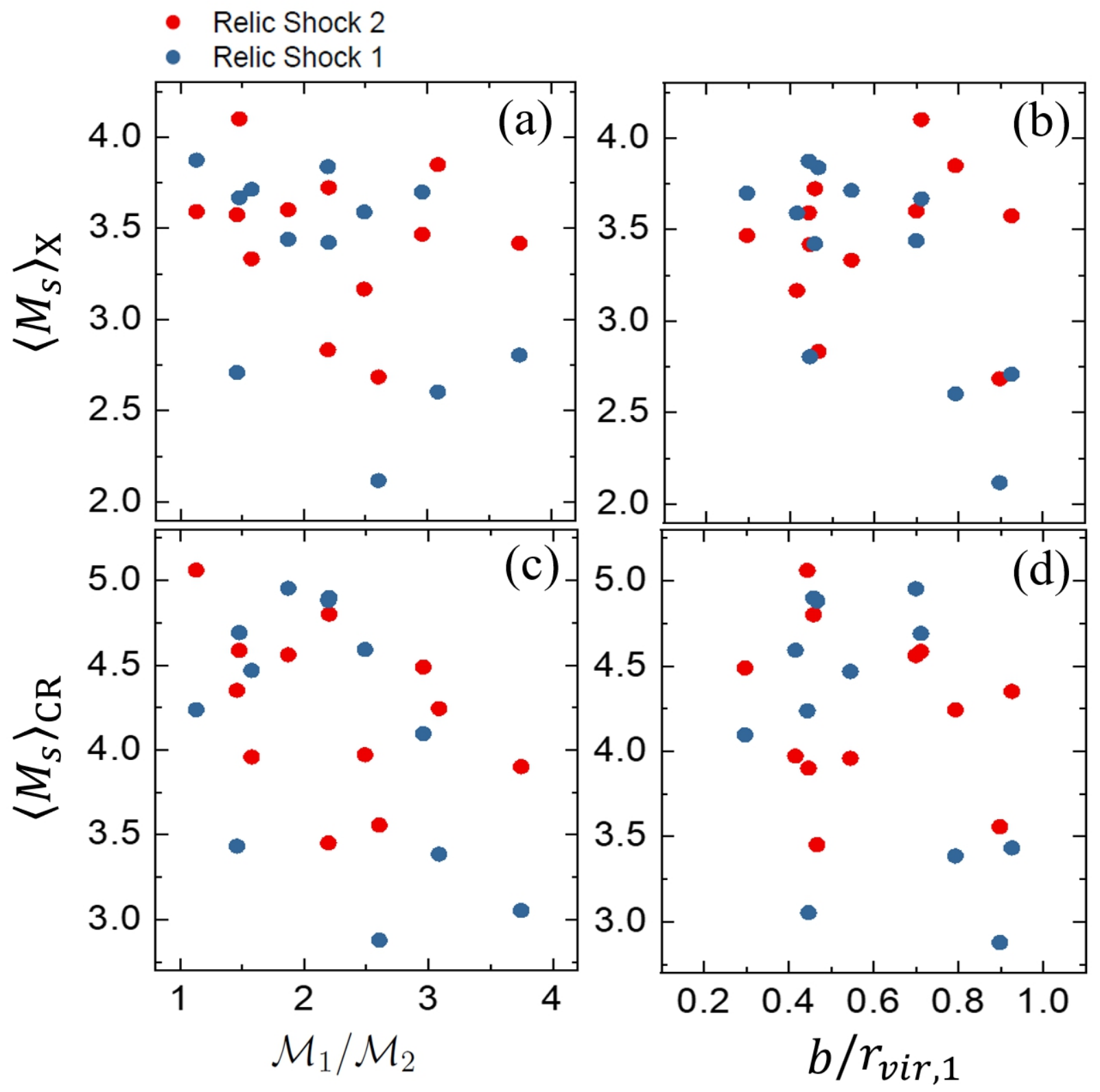}
\vskip -0.2 cm
\caption{${\langle M_s \rangle}_X$ (top panels) and ${\langle M_s \rangle}_{\rm CR}$ (bottom panels) versus mass ratio $\mathcal{M}_1/\mathcal{M}_2$ (left panels) and impact parameter $b/r_{\rm vir,1}$ (right panels) for 24 merger shocks in 12 sample clusters.}\label{f9}
\end{figure}

Figure \ref{f8}(b) presents ${\langle M_s \rangle}_{\rm X}$ and ${\langle M_s \rangle}_{\rm CR}$, both of which are linked to observable quantities, as a function of the normalized mean distance of merger shocks from the cluster center, $\left<d_s\right>/r_{\rm vir}$. They generally tend to increase with $\langle d_s \rangle/r_{\rm vir}$. This trend is partly due to the fact that, on average, the gas density decreases and the shock speed increases outwardly in merged clusters \citep{Ha2018}. There is substantial scatter, which would be again attributed to the merger histories of clusters and the influence of turbulent flow motions in the ICM.

The Mach number distribution of merger shocks, and hence the representative Mach numbers, could depend on merger parameters, similar to the morphological properties discussed in Section \ref{s3.1}. Figure \ref{f9} shows ${\langle M_s \rangle}_{\rm X}$ and ${\langle M_s \rangle}_{\rm CR}$ as functions of $\mathcal{M}_1/\mathcal{M}_2$ and $b/r_{\rm vir,1}$. The average Mach numbers tend to decrease with increasing $\mathcal{M}_1/\mathcal{M}2$, although there is relatively large scatter, but the dependence on $b/r_{\rm vir,1}$ appears to be only marginal. {In \citet{Ha2018}, where cases with $b\approx0$ and $\mathcal{M}_1/\mathcal{M}_2\approx 2$ were examined, relic shock 1 ahead of the more massive subcluster tends to be stronger with higher Mach numbers than relic shock 2, albeit with large fluctuations (see their Figure 5(a)). However, this tendency is not obvious in our merger shock samples from clusters with nonzero impact factors. As can be seen in Figures \ref{f9}(a) and (b), in only 5 of 12 clusters, ${\langle M_s \rangle}_{\rm X}$ for relic shock 1 is higher than for relic shock 2. Similarly, in Figures \ref{f9}(c) and (d), in only 6 of 12 clusters, ${\langle M_s \rangle}_{\rm CR}$ for relic shock 1 exceeds that for relic shock 2. Given the complex interplay of various factors such as merger parameters, merger history, and ICM turbulence, however, we argue that a larger cluster sample would be necessary to establish a more robust conclusion on the relative strengths of $M_s$ between relic shock 1 and 2.}

\begin{deluxetable*}{lccccccc}[t]\label{t2}
\tabletypesize{\scriptsize}
\tablecolumns{8}
\tablenum{2}
\tablewidth{0pt}
\tablecaption{Properties of Observed Merger shocks in Galaxy Clusters Hosting Double Radio Relics}
\tablehead{
\colhead{Cluster Name}&
\colhead{$z$}&
\colhead{$T_{\rm X}$}&
\colhead{$d_s$}&
\colhead{$\alpha_{\rm int}$}&
\colhead{$M_{\rm radio}$ }&
\colhead{$M_{\rm X-ray}$ } &
\colhead{$M_{\rm X-ray}$ }\\
\colhead{}&
\colhead{}&
\colhead{(keV)}&
\colhead{$(\rm Mpc)$}&
\colhead{}&
\colhead{${(\alpha_{\rm int})}$}&
\colhead{${(T_{\rm jump})}$}&
\colhead{${(\rho_{\rm jump})}$}}
\startdata
Abell 3376 (W) & 0.046 & 4.2 & 1.6 & $1.17^{+0.06}_{-0.06}$ & $3.57^{+0.58}_{-0.58}$ & $2.8^{+0.4}_{-0.4}$ & n/a\\
Abell 3376 (E) & 0.046 & 4.2 & 0.4 & $1.37^{+0.08}_{-0.08}$ & $2.53^{+0.23}_{-0.23}$ & $1.5^{+0.10}_{-0.10}$ & n/a\\
Abell 3667 (NW) & 0.056 & 6.0 & 1.5 & $1.4^{+0.1}_{-0.1}$ & $2.45^\#$ & $3.34^{+0.91}_{-0.50}$ &$2.05^{+0.73}_{-0.38}$\\
Abell 3667 (SE) & 0.056 & 6.0 & 1.1& $1.2^{+0.1}_{-0.1}$ & $3.32^\#$ & $1.8^{+0.5}_{-0.4}$ & $1.3^{+0.1}_{-0.1}$\\
Abell 3365 (E) & 0.093 & 3.1 & 1.0 & $0.85^{+0.03}_{-0.03}$ & * & $3.5^{+0.6}_{-0.6}$ & n/a\\
Abell 3365 (CW) & 0.093 & 3.1 & 1.0 & $0.76^{+0.08}_{-0.08}$ & * & $3.9^{+0.8}_{-0.8}$ & n/a\\
ZwCl 0008.8+5215 (W) & 0.104 & 4.8 & 0.7& $1.59^{+0.06}_{-0.06}$ & $2.1^\#$ & $2.35^{+0.74}_{-0.55}$ & $1.48^{+0.50}_{-0.32}$ \\
ZwCl 0008.8+5215 (E) & 0.104 & 4.8 & 0.9& $1.49^{+0.12}_{-0.12}$ & $2.25^\#$ & $1.54^{+0.65}_{-0.47}$ &n/a \\
Abell 3186/MCXC J0352.4-7401  (NW) & 0.127 & 5 & 1.5 & $1.0^{+0.1}_{-0.1}$  & * & n/a & n/a \\
Abell 3186/MCXC J0352.4-7401  (SE) & 0.127 & 5 & 1.2 & $0.9^{+0.1}_{-0.1}$ & * & n/a & n/a  \\
Abell 1240 (NW) & 0.159 & 6.0 & 0.7 & $1.08^{+0.05}_{-0.05}$ & $5.1^{+3.10}_{-1.10}$ & $1.57^{+0.34}_{-0.34}$ & 2.0 \\
Abell 1240 (SE) & 0.159 & 6.0 & 1.1 & $1.13^{+0.05}_{-0.05}$ & $4.0^{+1.10}_{-0.60}$ & $1.71^{+0.25}_{-0.25}$ & 2.0 \\
Abell 2345 (E) & 0.179 & $\sim 5 ^\dag$ & 0.9 & $1.29^{+0.07}_{-0.07}$ & $2.8^\#$ & n/a & n/a \\
Abell 2345 (W) & 0.179 & $\sim 5 ^\dag$ & 1.0 & $1.52^{+0.08}_{-0.08}$ & $2.2^\#$ & n/a & n/a \\
8C 0212+703/CIG 0217+70 (W) & 0.180 & 9.1 & 2.4& $1.01^{+0.05}_{-0.05}$ & 14 & n/a  & n/a \\
8C 0212+703/CIG 0217+70 (E) & 0.180 & 9.1 & 2.5& $1.09^{+0.06}_{-0.06}$ & $4.8^{+3.4}_{-1.0}$ & n/a & n/a\\
CIZA J2242.8+5301 (N) & 0.192 & 8.4 & 1.5& $1.12^{+0.03}_{-0.03}$ & $4.2^{+0.4}_{-0.6}$ & $2.7^{+0.7}_{-0.4}$ & n/a \\
CIZA J2242.8+5301 (S) & 0.192 & 8.4 & 1.1 & $1.12^{+0.07}_{-0.07}$ & $4.2^\#$ & $1.7^{+0.4}_{-0.3}$ & n/a \\
Abell 2146 (NW) & 0.232 & 6.75 & 0.45 &  $1.14^{+0.08}_{-0.08}$ & $3.91^\#$ & $2.0^{+0.3}_{-0.3}$ & $1.6^{+0.1}_{-0.1}$ \\
Abell 2146 (SE) & 0.232 & 6.75 & 0.2 & $1.25^{+0.07}_{-0.07}$&  $3.0^\#$ & $1.8^{+0.3}_{-0.2}$
& $2.3^{+0.2}_{-0.2}$  \\
RXC J1314.4-2515 (W) & 0.247 & 8.7 & 0.5 & $1.22^{+0.09}_{-0.09}$ & $3.18^{+0.87}_{-0.45}$ & $2.4^{+1.10}_{-0.8}$ & $1.7^{+0.4}_{-0.2}$ \\
RXC J1314.4-2515 (E) & 0.247 & 8.7 & 1.0 & $1.41^{+0.09}_{-0.09}$& $2.42^\#$ & n/a & n/a \\
ZwCl 2341.1+0000 (NW) & 0.270 & 5.3 & 0.8 & $1.02^{+0.02}_{-0.02}$&$10.05^\#$& n/a & $2.06^{+1.39}_{-0.76}$ \\
ZwCl 2341.1+0000 (SE) & 0.270 & 5.3 & 0.9 & $0.98^{+0.02}_{-0.02}$ &*& n/a & $1.43^{+0.23}_{-0.20}$\\
SPT-CL J2032-5627 (NW) &0.284 & 4.99 &1.2 & $1.18^{+0.10}_{-0.10}$ & $3.48^\#$ & n/a & n/a \\
SPT-CL J2032-5627 (SE) &0.284 & 4.99 & 0.4 & $1.52^{+0.10}_{-0.10}$ & $2.20^\#$ & n/a & n/a \\
PSZ1 G096.89+24.17/ZwCl 1856.8+6616 (N) & 0.3 & 3.7 & 0.75 & $0.95^{+0.07}_{-0.07}$ & * & n/a & n/a \\
PSZ1 G096.89+24.17/ZwCl 1856.8+6616 (S) & 0.3 & 3.7 & 1.1 & $1.17^{+0.07}_{-0.07}$ & $3.6^{+0.7}_{-0.7}$ & n/a & n/a \\
PSZ1 G108.18-11.53 (N) & 0.335 & $\sim 6.5^\dag$ & 1.7 & $1.25^{+0.02}_{-0.02}$ & $3.0^\#$ & n/a & n/a \\
PSZ1 G108.18-11.53 (S) & 0.335 & $\sim 6.5^\dag$ & 1.3 & $1.28^{+0.02}_{-0.02}$ & $2.85^\#$ & n/a & n/a \\
PSZ2 G233.68+36.14 (N) & 0.345 & 9.0& 0.7 & 
$1.31^{+0.12}_{-0.12}$ & $2.7^{+0.7}_{-0.4}$ & n/a & n/a
\\
PSZ2 G233.68+36.14 (SE) & 0.345 & 9.0 & 1.0 & $0.97^{+0.12}_{-0.12}$ &* & n/a & n/a \\
MACS J1752.0+4440 (NE) & 0.366 & 5.9 & 1.3& $1.16^{+0.03}_{-0.03}$ & $3.7^\#$& n/a & n/a  \\
MACS J1752.0+4440 (SW) & 0.366 & 5.9 & 0.8& $1.10^{+0.05}_{-0.05}$ & $4.6^\#$& n/a & n/a \\
ZwCl 1447.2+2619 (N) & 0.376 & 4.5 & 0.5 & $1.27^{+0.31}_{-0.31}$ & $2.9^{+0.8}_{-0.8}$ & n/a & n/a\\
ZwCl 1447.2+2619 (S) & 0.376 & 4.5 & 0.8 & $1.68^{+0.30}_{-0.30}$ & $2.0^{+0.7}_{-0.7}$& n/a & n/a\\
PLCK G287.0+32.9/PSZ2 G286.98+32.90 (NW) & 0.39 & 12.86 & 0.4 & $1.19^{+0.03}_{-0.03}$ & $3.4^\#$ & n/a & n/a\\
PLCK G287.0+32.9/PSZ2 G286.98+32.90 (SE) & 0.39 & 12.86 & 2.8 & $1.36^{+0.04}_{-0.04}$ & $2.56^\#$ & n/a & n/a \\
El Gordo/ACT-CLJ0102-4915 (NW) & 0.87 & 14.5 & 1.4 & $1.25^{+0.04}_{-0.04}$ & $3.0^{+0.2}_{-0.2}$ & $ 2.5 ^{+0.7}_{-0.3}$  & $2.78^{+0.63}_{-0.38}$\\
El Gordo/ACT-CLJ0102-4915 (E)  & 0.87 & 14.5  & 0.4 & $1.06\pm0.04$ & $6^{+2}_{-2}$ & n/a & n/a \\
\enddata 

\tablecomments{
Column1: Name of galaxy clusters with double radio relics.
Column2: Redshift of cluster.
Column3: X-ray temperature of cluster.
Column4: Projected distance of merge shock from the cluster center.
Column5: Integrated radio spectral index.
Column6: Radio Mach number calculated with $\alpha_{\rm int}$.
Column7: X-ray Mach number derived from temperature jump.
Column8: X-ray Mach number derived from density jump.\\
The \# symbol indicates that $M_{\rm radio}$ is calculated using $\alpha_{\rm int}$ given in the reference.
The * symbol means that $M_{\rm radio}$ can not be estimated, because $\alpha_{\rm int} < 1$.
The $\dag$ symbol indicates that $\left< kT_{\rm X}\right>$ is estimated using the $L_X-T$ relation in \citet{Pratt2009}.
For cases where shocks are not detected in X-rays or X-ray Mach numbers are not reported, n/a is indicated.}

\tablerefs{Abell 3376: \citet{Akamatsu2013}, \citet{George2015}, \citet{Urdampilleta2018}; Abell 3667: \citet{Akamatsu2013},  \citet{sarazin2016},  \citet{Storm2018}, \citet{deGasperin2022}; Abell 3365: \citet{Duchesne2021}, \citet{Urdampilleta2021}; ZwCl 0008.8+5215: \citet{vanWeeren2011b}, \citet{kierdorf2017}, \citet{Digennaro2019}; Abell 3186: \citet{Nesci1997}, \citet{Duchesne2021}; Abell 1240: \citet{Barrena2009}, \citet{Hoang2018}, \citet{sarkar2024}; Abell 2345: \citet{George2017}, \citet{Stuardi2021}; 8C 0212+703: \citet{Hoang2021}, \citet{Tumer2023}; CIZA J2242.8+5301: \citet{Akamatsu2013}, \citet{Akamatsu2013}, \citet{DiGennaro2018}, \citet{Loi2020}; Abell 2146: \citet{Russell2011}, \citet{Russell2012}, \citet{Hoang2019}; RXC J1314.4-2515: \citet{Valtchanov2002}, \citet{venturi2007}, \citet{George2017}, \citet{Stuardi2019}, \citet{Botteon2020}; ZwCl 2341.1+0000 : \citet{zhang2021}, \citet{Stuardi2022}; SPT-CL J2032-5627: \citet{Bulbul2019} \citet{Duchesne2021b} ; PSZ1 G096.89+24.17: \citet{Finner2021}, \citet{Jones2021}; PSZ1 G108.18-11.53: \citet{degasperin2015}; PSZ2 G233.68+36.14: \citet{Ghirardini2021}; MACS J1752.0+4440: \citet{vanWeeren2012}, \citet{Finner2021}; ZwCl 1447.2+2619: \citet{Lee2022}; PLCK G287.0+32.9: \citet{Bagchi2011}, \citet{George2017}; El Gordo: \citet{Menanteau2012}, \citet{Botteon2020}, \citet{Stuardi2022}}

\end{deluxetable*}

\section{Comparison with Observations} \label{s4}

In this section, we attempt to compare the properties of merger shocks reproduced in our simulations with those inferred from observed double radio relics. So far, about a hundred clusters hosting radio relics have been identified, and of those, approximately two dozen have double relics \citep[see, e.g.,][]{vanWeeren2019}. Table \ref{t2} provides a summary of the observational data and the Mach numbers inferred from radio and X-ray observations for the double radio relics reported in the literature \citep[see][and the references listed below the table]{Stuardi2022}. Here, $M_{\rm radio}$ is the radio Mach number calculated using the integrated radio spectral index, $\alpha_{\rm int}$, while the X-ray Mach number, $M_{\rm X-ray}$, is estimated from either the temperature jump, $T_{\rm jump}$, or the density jump, $\rho_{\rm jump}$, across the relic shock. And $d_s$ is the distance from the cluster center, projected onto the sky.

\begin{figure}[t]
\vskip 0.2cm
\hskip -0.9 cm
\centering
\includegraphics[width=1.1\linewidth]{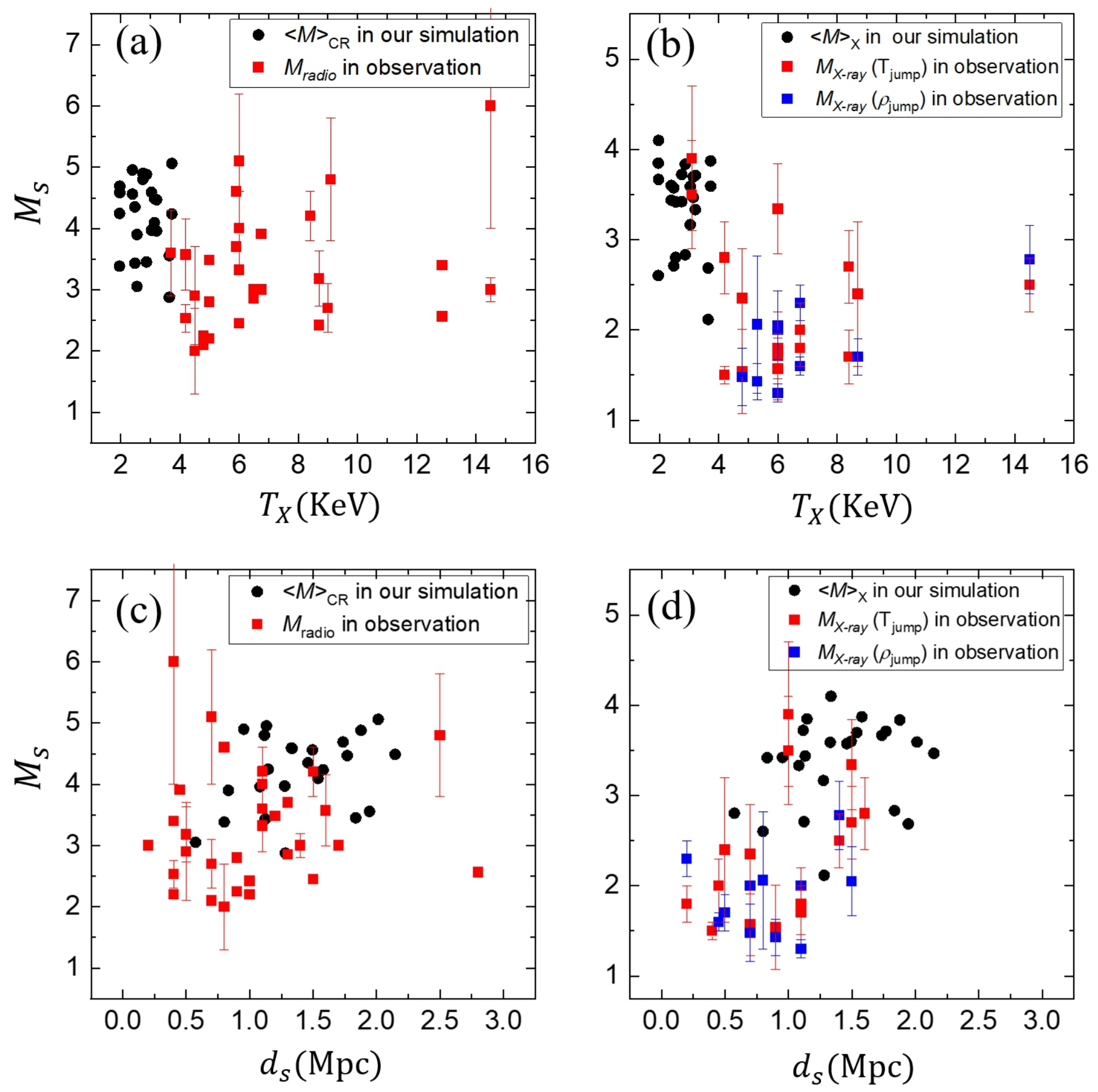}
\vskip -0.2 cm
\caption{(a) Comparison of ${\langle M_s \rangle}_{\rm CR}$ (black circles) of simulated merger shocks in our sample clusters with $M_{\rm radio}$ (red squares) of observed radio relics in Table \ref{t2}, plotted as a function of $T_X$. (b) Comparison of ${\langle M_s \rangle}_{\rm X}$ (black circles) of simulated merger shocks with $M_{\rm X-ray}$ (red and blue squares) of observed radio relics, plotted as a function of $T_X$. (c) and (d) Plots similar to (a) and (b), but shown as a function of the radio relic's distance from the cluster center, $d_s$. Here, for simulated radio relics, $d_s$ is the mean 3D distance of the shock surface, $\langle d_s \rangle$. In contrast, for observed radio relics, $d_s$ refers to the distance "projected" onto the sky, listed in Table \ref{t2}. The vertical lines denotes the errors in observation.}\label{f10}
\end{figure}

\begin{figure}[t]
\vskip 0.2cm
\hskip -0.3 cm
\centering
\includegraphics[width=1.0\linewidth]{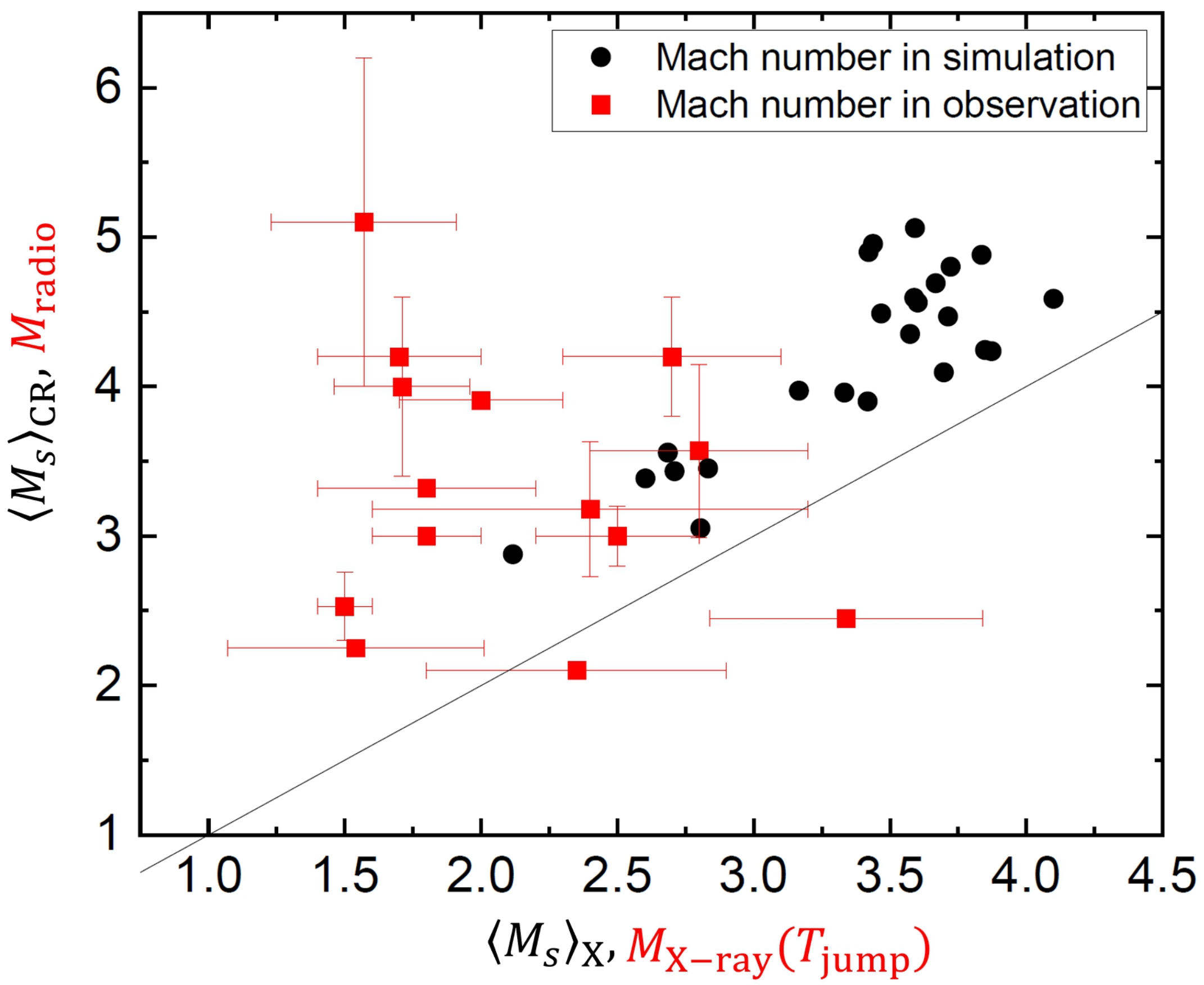}
\vskip -0.2 cm
\caption{${\langle M_s \rangle}_{\rm CR}$ versus ${\langle M_s \rangle}_{\rm X}$ (black circles) for simulated relic shocks in our sample clusters, and $M_{\rm radio}$ versus $M_{\rm X-ray}(T_{\rm jump})$ (red square) for observed radio relics in Table \ref{t2}. The vertical and horizontal lines denotes the errors in observation. The tilted line draws $x=y$.}\label{f11}
\end{figure}

In Figure \ref{f10}, we plot ${\langle M_s \rangle}_{\rm CR}$ and ${\langle M_s \rangle}_{\rm X}$ as functions of the X-ray temperature, $T_X$, and the average distance from the center, $\langle d_s \rangle$, for our 24 simulated merger shocks, along with $M_{\rm radio}$ and $M_{\rm X-ray}$ as functions of $T_X$ and the "projected" distance, $d_s$, for the observed radio relics listed in Table \ref{t2}. We compare ${\langle M_s \rangle}_{\rm CR}$ with $M_{\rm radio}$, assuming that ${\langle M_s \rangle}_{\rm CR}$ serves as a proxy of radio-emissivity-weighted average Mach number, since radio synchrotron emissions are produced by CR electrons accelerated via diffusive shock acceleration (DSA). Similarly, we compare ${\langle M_s \rangle}_{\rm X}$ with $M_{\rm X-ray}$, noting that ${\langle M_s \rangle}_{\rm X}$ is derived from 3D shock properties in our simulations, whereas $M_{\rm X-ray}$ is estimated from the observed X-ray surface brightness map projected onto the sky.

A couple of points are notable. First, Figures \ref{f10}(a) and (b) show that $T_X$ for simulated merger shocks (filled black circles) is, on average, lower than $T_X$ for observed radio relics (filled red and blue squares). This discrepancy arises because the merged clusters in our simulations have relatively low $T_X\lesssim 4$ keV due to the limited volume of our simulation box, as pointed in Section \ref{s2.2}. In contract, X-ray and radio observations preferentially pick up clusters with heavier mass and higher $T_X$. Hence, we present results in dimensionless quantities that would be roughly scale-independent, such as $M_s$ and $\left<d\right>/r_{\rm vir}$. Second, Figures \ref{f10}(c) and (d) indicate that $\langle d_s \rangle$ for simulated merger shocks is, on average, larger than $d_s$ for observed radio relics. This should be partly because the distances projected onto the sky are less than the true 3D distances for observed radio relics. We point out that $\langle d_s \rangle\sim r_{\rm vir}$ for simulated merger shocks, although there is considerable  scatter, as shown in Figure \ref{f4}(b); radio relics are also observed around $r_{\rm vir}$, as discussed in the introduction.

Given the differences between simulated and observed quantities, the comparison should proceed with caution. Nevertheless, we point out the following: (1) The simulations predict ${\langle M_s \rangle}_{\rm CR} \sim 3-5$ for merger shocks, whereas the observed relics exhibit $M_{\rm radio} \sim 2-5$. Considering the limitations inherent in our simulations, this level of agreement would be considered reasonable. On the other hand, it was argued that the critical Mach number for electron preacceleration and subsequent injection to DSA is $M_{\rm crit}\approx2.3$ \citep[e.g.,][]{Kang2019,Ha2021,Ha2022}. Several observed relics have low $M_{\rm radio}$ close to $\sim 2$; however, these relics likely have regions with $M_s \gtrsim M_{\rm crit}$ where electrons DSA may occur. (2) In contrast, ${\langle M_s \rangle}_{\rm X} \sim 2-4$ for simulated merger shocks looks greater than $M_{\rm X-ray} \sim 1.2-4$ for observed relics in Figure \ref{f10}. This discrepancy may partly arise from projection effects in X-ray observations \citep{Hong2015,Hoang2018}. Moreover, \citet{Wittor2021} demonstrated that $M_{\rm X-ray}$ derived from the X-ray surface brightness map projected onto the sky plane can vary significantly depending on the relic's orientation.

In Figure \ref{f11} we show the relation between the radio and X-ray Mach numbers for both simulated merger shocks and observed radio relics. In both samples, $M_{\rm radio} \gtrsim M_{\rm X-ray}$, confirming the earlier interpretation in Section \ref{s3.4} that electron acceleration is more efficient in portions of shock surfaces with higher $M_s$, and consequently, $M_{\rm radio}$ tends to be more heavily weighted by those higher $M_s$ regions. However, there are exceptions, such as Abell 3667 (NW) and ZwCl 0008.8+5215 (W), where $M_{\rm radio} \lesssim M_{\rm X-ray}$; these cases require more careful interpretation of radio and X-ray Mach numbers.

\section{Summary}\label{s5}

A major merger between two subclusters of comparable masses is expected to produce a pair of megaparsec-scale bow shocks that propagate into the outskirts of the merged cluster over gigayears following pericenter passage \citep[e.g.,][]{vanWeeren2011a}. When projected onto the sky, these merger shocks often appear as double radio relics in radio observations or as temperature and surface brightness discontinuities in X-ray observations (see Table \ref{t2}).

In this paper, we analyzed the data from a set of cosmological hydrodynamic simulations to investigate the properties of such shocks induced by major binary mergers. We focused on 12 merging clusters with mass ratio $1\!\le \! \mathcal{M}_1/\mathcal{M}_2 \! \lesssim \!4$ and normalized impact parameter $b/r_{\rm,1}\! \lesssim \! 1$, resulting in a total of 24 merger shocks examined at the optimal redshift for radio relic observation, $z_{\rm relic}$ (see Table \ref{t1}). Relic shock 1 forms in front of the heavier subcluster 1, while relic shock 2 forms in front of the lighter subcluster 2.
We isolated the shock zones associated with the shock surfaces, distinguishing them from background ICM shocks produced by turbulence and infall, based on the criteria outlined in Section \ref{s2.3}. 

We inspected the morphological characteristics of these merger shock surfaces, including their shape and area. We then quantified the Mach number distribution across the shock surfaces. Since merger shocks consist of numerous zones with varying Mach numbers, we defined representative average Mach numbers. We explored how these representative Mach numbers manifest in simulated merger shocks and compared them with the Mach numbers inferred in X-ray and radio observations.

Our main results are summarized as follows:

1. In our sample, merger shock surfaces are located at $\langle d_s \rangle / r_{\rm vir}\! \approx\! 0.6 - 1.2$ from the X-ray center of the clusters at $z_{\rm relic}$ (see Figure \ref{f4}). The shock surfaces can be approximated as elliptical sections of spherical shells, with an axial ratio of $b_{\rm ss}/a_{\rm ss} \gtrsim 0.6$. Each shock surface covers approximately $\sim5-20\ \%$ of the surface area of the shock sphere with $r_{\rm sphere} \!=\! \langle d_s \rangle$. The surface area ratio of relic shock 2 to relic shock 1 scales roughly with the mass ratio of subclusters as $A_{\rm ss,2}/A_{\rm ss,1} \propto \mathcal{M}_1/\mathcal{M}_2$ (see Figure \ref{f5}).

2. Due to nonzero impact parameters and turbulent ICM flows, merger shock surfaces become distorted, with their centers offset from the axis connecting the two DM density peaks (see Figures \ref{f6}(a) and \ref{fA2}). As a result, the distribution of the average Mach number of shock zones, as a function of the position angle $\theta$ between the axis connecting the two DM density peaks and the line extending from the mid-point of these peaks to the shock zone, deviates from the typical $\cos \theta$ pattern. Instead, these distributions often peak at fairly large $\theta$, or exhibit significant fluctuations without a distinct peak (see Figures \ref{f6}(b) and \ref{fA3}).

3. The PDFs for $M_s$ of shock zones across merger shock surfaces are positively skewed and well-fitted by a log-normal function (see Figures \ref{f7}(a) and \ref{fA4}). The fitted log-normal distributions peak around $M_{s,\rm peak} \approx 2-4.5$ with tails extending up to $M_s\sim10$. The ratio of the standard deviation to the mean of the PDFs is approximately $\sigma_{M_s}/\langle M_s \rangle_{\rm area}\approx 0.37$ (see Figures \ref{f7}(b)). Here, the mean of the PDF is the shock-surface-area-weighted average Mach number, $\langle M_{s} \rangle_{\rm area}$.

4. The area-weighted average Mach numbers, ${\langle M_s \rangle}_{\rm area} \! \approx \!2.3-4.4$ are comparable to the X-ray-emissivity-weighted average Mach numbers, ${\langle M_s \rangle}_{\rm X}\! \approx\! 2-4$. In contrast, the CR-energy-flux-weighted average Mach numbers are higher with ${\langle M_s \rangle}_{\rm CR}\! \approx\! 3-5$ (see Figure \ref{f8}). This discrepancy aligns with observations of radio relics, where generally the Mach numbers inferred from radio data are greater than those derived from X-ray data, $M_{\rm radio} \gtrsim M_{\rm X-ray}$.

5. In our simulated merger shocks, both ${\langle M_s \rangle}_{\rm X}$ and ${\langle M_s \rangle}_{\rm CR}$ exhibit a weak, decreasing trend with increasing $\mathcal{M}_1/\mathcal{M}_2$, albeit with significant scatter. On the contrary, the dependence of these Mach numbers on $b/r_{\rm vir,1}$ appears to be only marginal (see Figure \ref{f9}).

6. For simulated merger shocks, we find ${\langle{M_s}\rangle}_{X}\gtrsim2$, while X-ray observations frequently report shocks with $M_{\rm X-ray}\lesssim2$ (see Figure \ref{f11}). {As suggested by several previous papers \citep[e.g.,][]{Hong2015,Wittor2021},} this discrepancy may be partly attributed to projection effects in X-ray observations, which can complicate accurate estimation of the true Mach numbers.

In this paper, we focused on major binary mergers with $1 \! \le\! \mathcal{M}_1/\mathcal{M}_2 \!\lesssim \! 4$ and $b/r_{\rm,1}\! \lesssim \! 1$, which are expected to present a relatively simple merger scenario. However, due to the complex interplay of factors such as merger history and turbulence, in addition to merger parameters, the structure of merger shocks turns out to be fairly complicated. 
Our aim was to characterize the properties of shock surfaces, yet a larger sample of simulated merging clusters appears necessary to establish more definitive relationships, such as the dependence of average Mach numbers on merger parameters.

Finally, some observed radio relics appear to have formed under merger conditions different from those considered here and also may result from multiple merger events. We leave investigations of these cases for future work.

\begin{acknowledgments}
{The authors would like to thank the anonymous referee for constructive comments and suggestions.} This work was supported by the National Research Foundation (NRF) of Korea through grants 2020R1A2C2102800, 2023R1A2C1003131, RS-2022-00197685, and 2023K2A9A1A01099513. We thank Dr. J.-H. Ha for discussions.
\end{acknowledgments}

\bibliography{reference}{}
\bibliographystyle{aasjournal}

\appendix
\section{Figures of All Sample Clusters}
\label{sec:sa}

\restartappendixnumbering{}

In the main part, we present figures for Clusters 1, 3, 6, and 11 to maintain conciseness. In this appendix, we provide figures for all 12 sample clusters for comprehensive reference.

\begin{figure*}[b]
\vskip 0 cm
\hskip 2.3 cm
\includegraphics[width=0.72\linewidth]{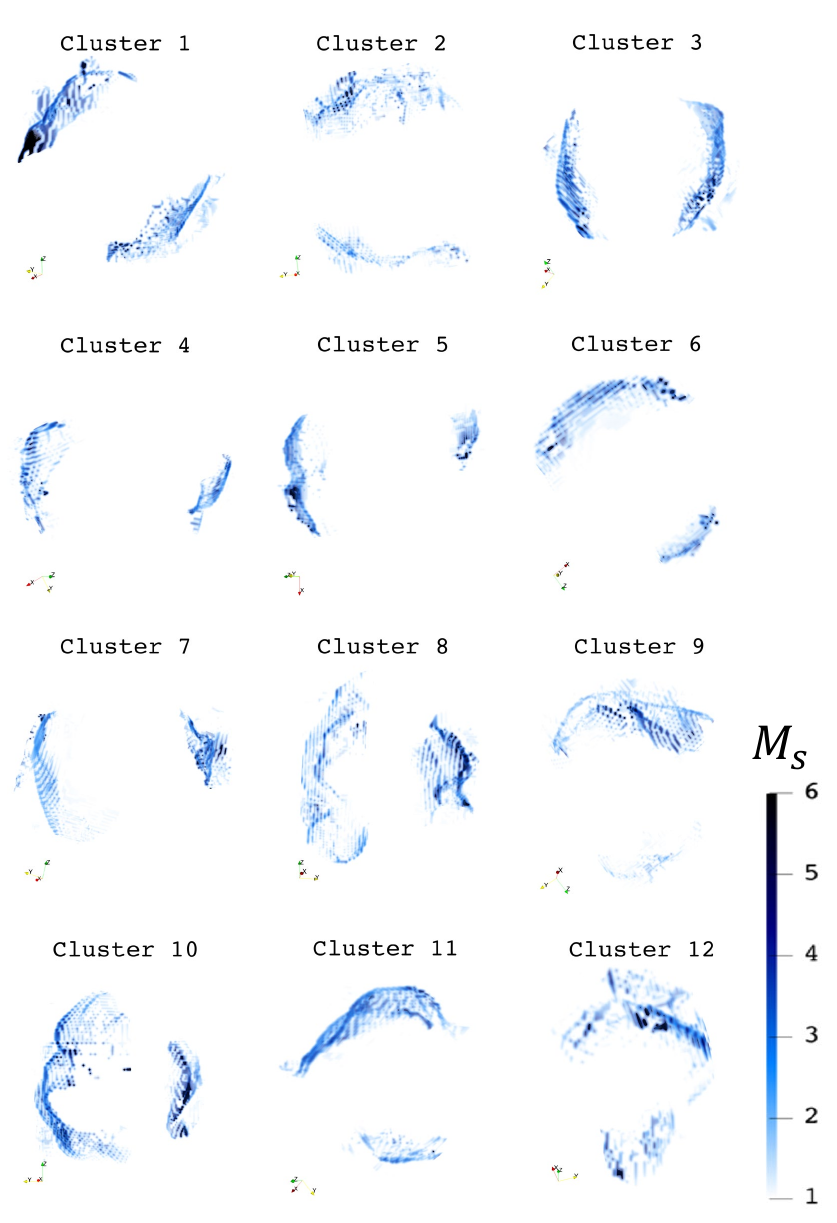}
\vskip -0.5 cm
\caption{Same as in Figure \ref{f4}(a) for all 12 sample clusters.}\label{fA1}
\end{figure*}

\begin{figure*}[t]
\vskip 0 cm
\hskip 2 cm
\includegraphics[width=0.75\linewidth]{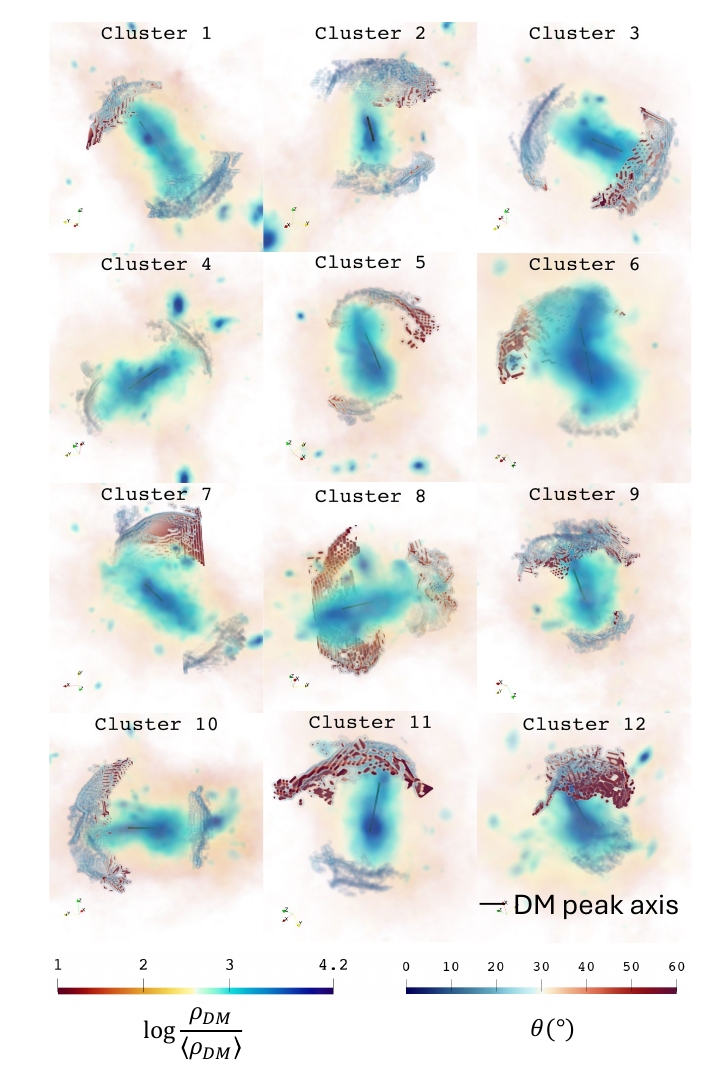}
\vskip -0.2 cm
\caption{Same as in Figure \ref{f6}(a) for all 12 sample clusters.}\label{fA2}
\end{figure*}

\begin{figure*}[t]
\vskip 0.1 cm
\hskip -0.1 cm
\includegraphics[width=1.0\linewidth]{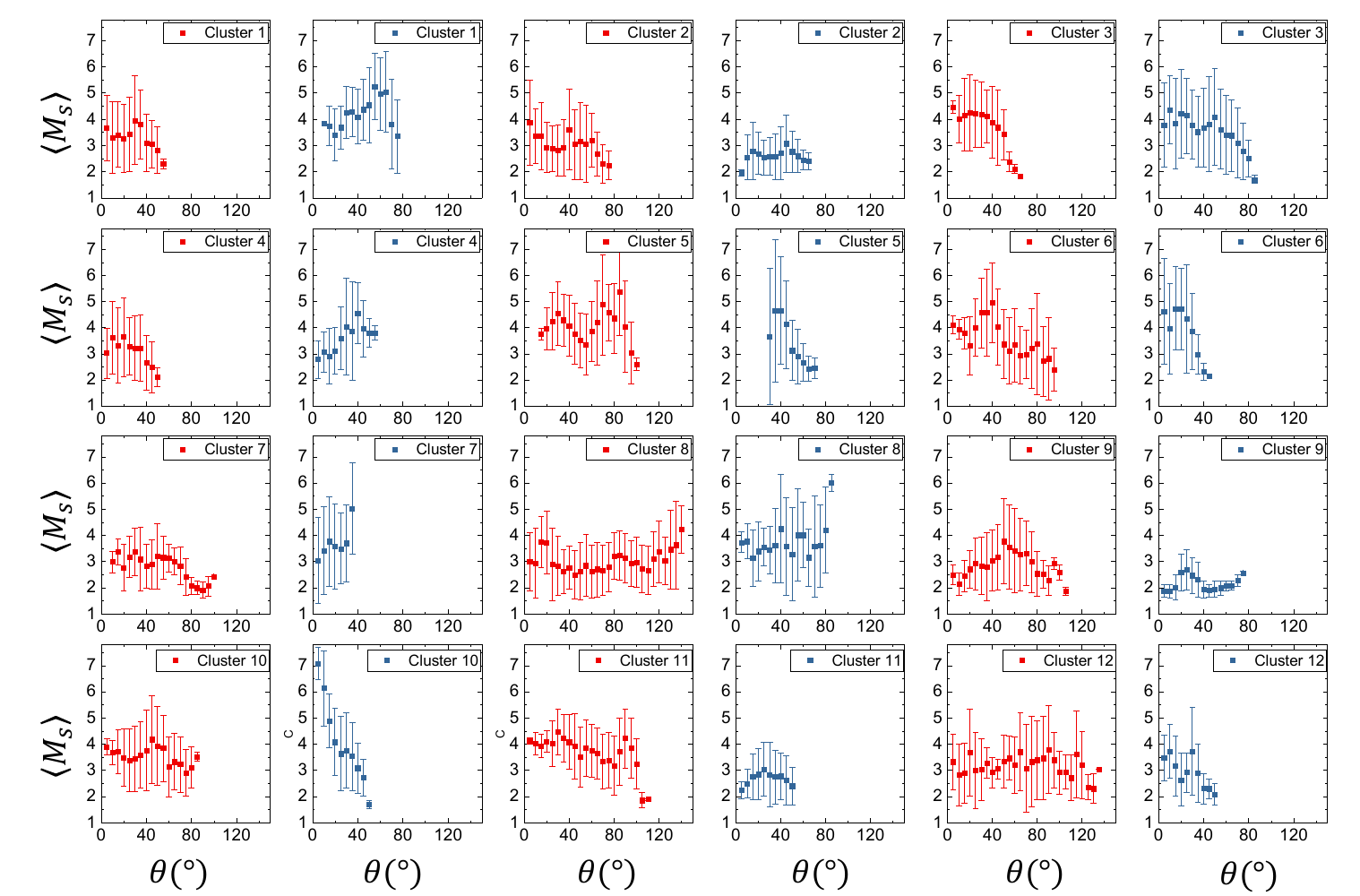}
\vskip -0.2 cm
\caption{Same as in Figure \ref{f6}(b) for all 12 sample clusters.}\label{fA3}
\end{figure*}

\begin{figure*}[t]
\vskip 0.1 cm
\hskip -0.1 cm
\includegraphics[width=1.0\linewidth]{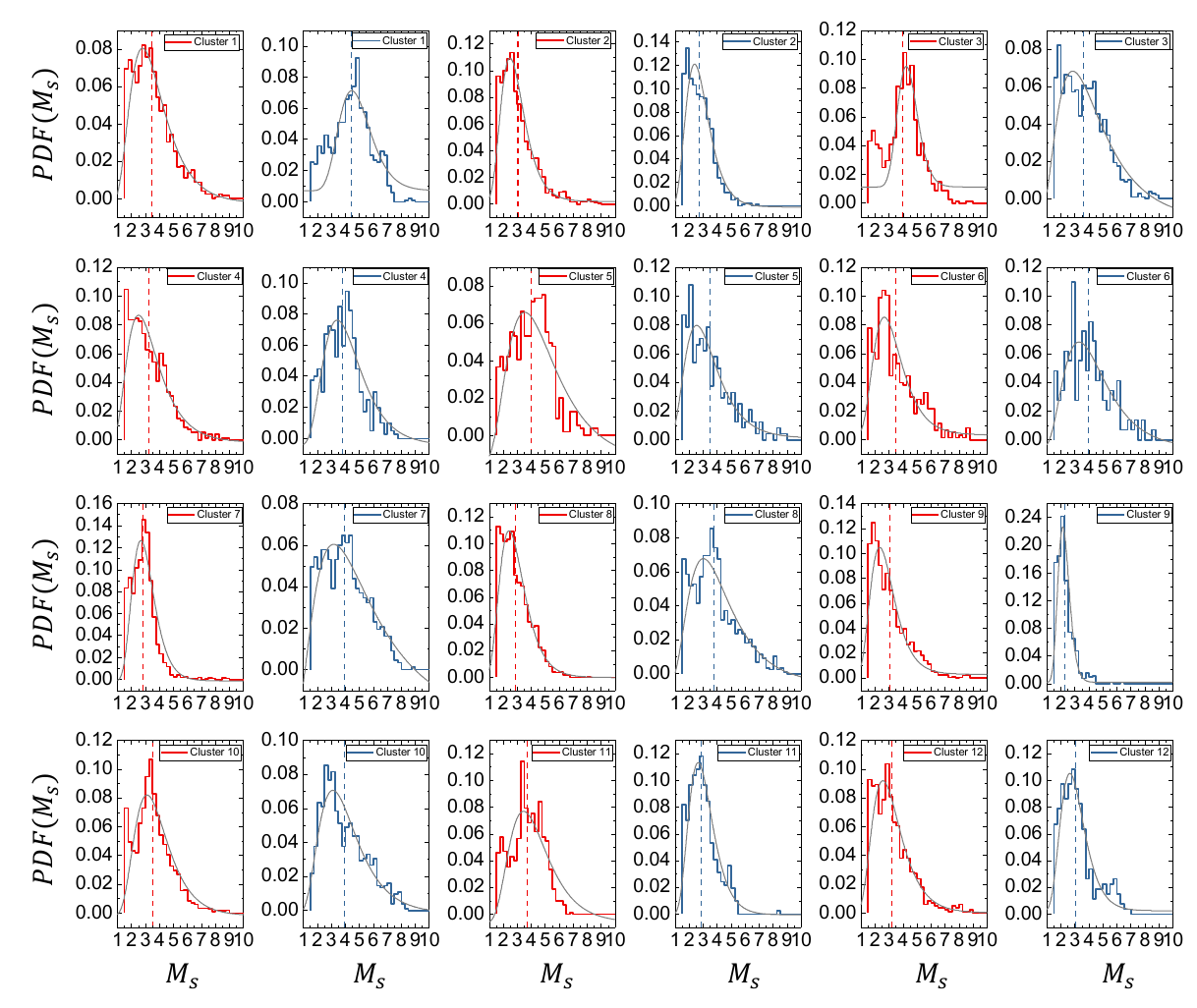}
\vskip -0.2 cm
\caption{Same as in Figure \ref{f7}(a) for all 12 sample clusters.}\label{fA4}
\end{figure*}

\end{document}